\documentclass[
 reprint,
superscriptaddress,
 amsmath,amssymb,
 aps,
pra,
onecolumn,
]{revtex4-2}
\usepackage{cancel}
\usepackage{enumerate}
\usepackage{graphicx}
\usepackage{dcolumn}
\usepackage{bm}
\usepackage{hyperref}

\hypersetup{pdfencoding=auto,colorlinks=true,linkcolor=blue,citecolor=blue}
\expandafter\let\csname equation*\endcsname=\relax 
\expandafter\let\csname endequation*\endcsname=\relax 
\usepackage{amsmath,amssymb}
\usepackage{dsfont}
\usepackage{braket}

\usepackage[english]{babel}
\usepackage[T1]{fontenc}

\newcommand{\argmax}[1]{\mathrm{arg}\max\limits_{#1}}
\graphicspath{{./}{./Figures/}}

\begin{document}

\title{A Monte Carlo Tree Search approach to QAOA: finding a needle in the haystack}

\author{Andoni Agirre}
\email{andoni.agirre@dipc.org}
\affiliation{Donostia International Physics Center (DIPC),  Manuel Lardizabal Pasealekua 4, 20018 Donostia, Basque Country} 
\affiliation{%
Center for Quantum Devices, Niels Bohr Institute, Copenhagen University, Universitetsparken 5, Copenhagen 2100, Denmark}
\affiliation{%
Department of PMAS: Physics, Chemistry and Technology, University of the Basque Country (UPV/EHU), Manuel  Lardizabal Pasealekua 3, 20018 Donostia, Basque Country} 

\author{Evert van Nieuwenburg}
 \affiliation{
 $\langle aQa \rangle$ at Lorentz Institute and Leiden Institute of Advanced Computer Science,
Leiden University, P.O. Box 9506, 2300 RA Leiden, The Netherlands
}%

\author{Matteo M. Wauter}
\email{matteo.wauters@unint.it}
\affiliation{CNR-INO Pitaevskii BEC Center and Department of Physics, University of Trento, Via Sommarive 14, I-38123 Povo, Trento, Italy}
\affiliation{INFN-TIFPA, Trento Institute for Fundamental Physics and Applications, Via Sommarive 14, I-38123 Povo, Trento, Italy}
\affiliation{%
Center for Quantum Devices, Niels Bohr Institute, Copenhagen University, Universitetsparken 5, Copenhagen 2100, Denmark}

\begin{abstract}
The search for quantum algorithms to tackle classical combinatorial optimization problems has long been one of the most attractive yet challenging research topics in quantum computing. 
In this context, variational quantum algorithms (VQA) are a promising family of hybrid quantum-classical methods tailored to cope with the limited capability of near-term quantum hardware. 
However, their effectiveness is hampered by the complexity of the classical parameter optimization which is prone to getting stuck either in local minima or in flat regions of the cost-function landscape.
The clever design of efficient optimization methods is therefore of fundamental importance for fully leveraging the potential of VQAs.
In this work, we approach QAOA parameter optimization as a sequential decision-making problem and tackle it with an adaptation of Monte Carlo Tree Search (MCTS), a powerful artificial intelligence technique designed for efficiently exploring complex decision graphs.
We show that leveraging regular parameter patterns deeply affects the decision-tree structure and allows for a flexible and noise-resilient optimization strategy suitable for near-term quantum devices.
Our results shed further light on the interplay between artificial intelligence and quantum information and provide a valuable addition to the toolkit of variational quantum circuits.
\end{abstract}

\maketitle

\section{\label{sec:level1}Introduction}
Overcoming the limitations of classical computing for problems hard to solve on classical machines, typically those requiring exponentially large resources, has been one of the main drives behind the development of quantum computation since the first pioneering works on quantum algorithms.
Indeed, some celebrated examples, such as Grover's algorithm for unstructured search~\cite{grover1996fast}, its generalization to more complex problems such as amplitude estimation~\cite{Brassard_2002} and minimum finding~\cite{Durr_1999}, and Shor's algorithm for factorization~\cite{shor_factoring} display significant speedup over classical methods, motivating further research in the quest for quantum advantage. 
However, the realization of fault-tolerant quantum computers, a prerequisite for fully exploiting this potential, remains a formidable challenge. 
In the interim, leveraging available technology necessitates exploring alternative strategies, such as variational quantum algorithms~\cite{McClean_NewJPhys2016, Moll_2018, Cerezo_NatRev2021, Bharti_NISQ} (VQA), to harness the computational power of current quantum devices.

VQAs employ a hybrid approach, combining classical optimization loops with (shallow) parameterized quantum circuits, offering a practical means of quantum computation with existing hardware. 
In the last years, they have been at the center of intense research activities and their proposals range from quantum optimization~\cite{Moll_2018, McClean_PRXQ2021} and ground state preparation~\cite{Peruzzo_NatComm14, mbeng2019quantum, Cerezo_npj2022} to quantum compiling~\cite{Khatri2019quantumassisted,Sharma_njp2020} and quantum machine learning~\cite{Biamonte_Nature2017,farhi2018classification,Havlícek_Nature2019}.
Designing efficient methods capable of dealing with hardware noise, for the classical optimization of the variational parameters, is crucial for their success. In this work, we propose a framework inspired by artificial intelligence techniques displaying remarkable resilience against noise.

One of the most used algorithms in the VQA family is the quantum approximate optimization algorithm~\cite{Farhi_arXiv2014,lloyd2018quantum, streif2019comparison, mbeng2019quantum, Zhou_PRX2020, Wauters_PRA2020, blekos2023review, Santra_PRA2024} (QAOA), which aims at finding an (approximate) solution to a classical combinatorial optimization problem mapped onto a cost Hamiltonian $H_C$, diagonal in the computational basis.
To find its ground state, which minimizes the energy and thus also the classical cost function, QAOA starts from the ground state of an auxiliary mixing Hamiltonian $H_M$. 
Typically, one chooses $H_M = \sum_i \sigma^x_i$, as its ground state $|+\rangle^{\otimes n} $ is easy to prepare and has a uniform weight on each state in the computational basis.
Then, QAOA prepares the variational state 
\begin{equation}
     \ket{\bm{\gamma}, \bm{\beta}} = \prod_{i=1}^P e^{-i\beta_i H_M}e^{-i\gamma_i H_C}\ket{+}^{\otimes n} \ ,
\end{equation}
where $P$ sets the circuit depth and $(\bm{\gamma}, \bm{\beta})$ is the array of variational parameters. Their optimal value is found by minimizing $F_P(\bm{\gamma},\bm{\beta})=\bra{\bm{\gamma}, \bm{\beta}} H_C \ket{\bm{\gamma}, \bm{\beta}}$.
QAOA has been applied to a wide range of problems, both classical and quantum, and many approaches have been proposed to improve its original formulation~\cite{blekos2023review}. 
Examples include several heuristics to improve the parameter optimization~\cite{wilson2019optimizing, Nakanishi_PRR2020, Zhou_PRX2020,Sack_TQA_init_QAOA, Lumia_PRXQuantum2022}, reinforcement-learning guided exploration of the energy landscape~\cite{yao2020policy, Wauters_PRR2020}, and addition of further unitary operations to make the ansatz more expressive~\cite{Hadfield_Algs2019, Yao_PRX2021, Zhu_PRR2022, Chandarana_PRR2022}. 

Despite the flexibility and popularity of VQAs, one of their main weaknesses appears in the classical optimization loop~\cite{BIttel_PRL2021}, which may suffer from hardware and measurement noise or the presence of flat regions in the cost function, known as barren plateaus~\cite{McClean_QAOA_barren_plateaus}. 
Indeed, both effects either slow down the exploration of parameter space or prevent it from escaping local minima,
in particular when the classical optimization algorithm relies on the gradient of the cost function.
To tame these issues,  gradient-free approaches have been suggested as appealing alternatives for the practical implementation of VQAs.
 However, even local minimization methods based on finite differences (also known as pseudo-gradient methods) suffer from the exponential scaling of resources when facing barren plateaus~\cite{Arrasmith_Quantum2021}.
True gradient-free global optimizers are promising candidates to overcome such difficulties.
 Among these, reinforcement-learning~\cite{Sutton_2ed} inspired approaches emerged as intriguing choices, both for the parameter optimization~\cite{yao2020policy, Wauters_PRR2020, chen_mctsQA} and circuit design~\cite{Yao_PRX2021, MCTS_QAOA_circuit}.

In this context, Monte Carlo Tree Search~\cite{Kocsis_MCTS} (MCTS) appears as a compelling gradient-free and robust optimization method for variational quantum algorithms in the presence of noise. 
Being the backbone of many recent achievements in machine learning of complex combinatorial games, most notably chess and Go \cite{alphazero_original}, MCTS has gained substantial fame and has been widely used for many different tasks across many different fields \cite{MCTS_review, mcts_survey}. 
As such, MCTS has already been successfully applied in the context of quantum annealing schedule optimization~\cite{chen_mctsQA, Wauters_2022}, control problems~\cite{Dalgaard2020} and circuit design~\cite{MCTS_QAOA_circuit}. 
Its problem-agnostic nature, the ease with which it can be integrated into deep learning frameworks, and, fundamentally, the different approach to the optimization task compared to more traditional algorithms, make it a promising candidate for enhancing the performance of variational quantum algorithms with complex parameter landscapes, such as QAOA. 

In this work, we investigate the potential of MCTS in identifying optimal strategies for QAOA applied to instances of the optimization versions of 3-SAT and MaxCut, both $\mathsf{NP}$-hard classical optimization problems \cite{Karp1972}. 
We show that a suitably modified version of MCTS offers a compelling practical route to design robust optimal QAOA schedules, with exceptional resilience against noise in the cost function.
The crucial modification to MCTS aims at exploiting the parameter concentration and regularity inherent in many optimal QAOA schedules~\cite{brandao2018fixed,Akshay_PRA2021, Streif_QST2020, Sack_TQA_init_QAOA}. 
This version, which we denote as iterative Search-Space Restricted Monte Carlo Tree Search (SSR-MCTS), progressively builds the solutions at each depth $P$ from the previous optima, inspired by iterative methods presented in Refs.~\cite{mbeng2019quantum, Zhou_PRX2020}. 
On the opposite, we find that a vanilla version of MCTS spectacularly fails in navigating the QAOA energy landscape already at a small circuit depth. 
This happens because of the particular fractal-like structure of the leaf nodes, the endpoints of the decision tree, where good strategies for QAOA are uniformly and sparsely distributed among bad ones. 
Iterative SSR-MCTS drastically simplifies the leaf-node structure of the QAOA decision tree, allowing MCTS to reliably find optimal schedules for different choices of the system size, circuit depth, or problem instance.
The combination of noise resilience, gradient-free optimization and good scalability in the quantum circuit volume makes our results well-suited for applications in current quantum hardware.

Our work showcases the emergence of optimization complexity in VQAs from the unusual perspective of sequential decision processes. Understanding the connection with other complexity estimators for the energy landscapes of QAOA can further unravel the interplay between artificial intelligence methods and quantum optimization, clarifying when the former provides valuable resources for NISQ devices.

The rest of the paper is organized as follows. 
In Section~\ref{sec:mcts} we recap the general structure of MCTS algorithms (\ref{sec:mcts_gen}), describe the MCTS approach for QAOA (\ref{sec:MCTS_for_QAOA}) and explain the SSR variant and its iterative version (\ref{sec:SSR}).
In sections \ref{sec:3sat} and \ref{sec:maxcut} we present our results on instances of 3-SAT and unweighted MaxCut on 3-regular graphs, respectively. We draw our conclusion and present future outlooks in Sec.~\ref{sec:conclusion}.
Further results and analysis are presented in the appendices: in Appendix~\ref{app:globalopt} we briefly discuss a combination of MCTS and gradient descent to construct a global optimization algorithm, in Appendix~\ref{app:ssrtree} we look at an example MCTS decision tree for QAOA, in Appendix~\ref{app:noise} we test the performance of various iterative SSR-MCTS variants on noisy cost landscapes, and in Appendix~\ref{app:cycles} we study the resource cost of our algorithm.

\section{MCTS-guided QAOA optimization}\label{sec:mcts}
In this section, we first give a brief introduction to the main concepts behind MCTS and its core elements. 
Then, we frame QAOA as a decision process and investigate the peculiar structure that makes it unsuitable for standard MCTS exploration. 
With this motivation, we describe in detail the modification needed to simplify the tree structure and optimize the policy for tackling QAOA.
A reader already familiar with MCTS can skip directly to Sec.~\ref{sec:MCTS_for_QAOA}.
\subsection{Monte Carlo Tree Search}\label{sec:mcts_gen}
\subsubsection*{General concepts.}

A decision tree is a structure that represents the possible paths originating from a sequential set of discrete action choices.
Each branching point ({\em node}) of the tree represents a set of decisions taken insofar, while the {\em children} of that node are the possible choices stemming from it.
We denote the starting point of the tree as the {\em root} node and the terminal states as {\em leaf} nodes.
Each leaf node has an associated reward depending on the success of that particular strategy.
As decision trees grow exponentially with their depth, clever algorithms are necessary to navigate such structures.

MCTS explores a given tree domain 
via random samples guided by some specified selection and playout policies, to find a path down the decision tree that maximizes a reward associated with the task at hand~\cite{Kocsis_MCTS, MCTS_review}.
Starting from the root node, it incrementally builds an asymmetric replica of the parts of the tree ({\em branches}) it deems most promising.
The algorithm repeatedly runs a four-stage cycle, learning about the tree by growing the part stored in memory. At the same time, it keeps track and continuously updates the cumulative score $w_i$ of each stored node $i$ and the number of times it has been visited by the algorithm $n_i$.
The cycles stop when the allotted computational budget is exceeded, at which point the algorithm makes a decision and suggests a strategy to follow. 
This usually corresponds either to the child node with the maximum average score $w_i/n_i$, or the one with the highest visit counts $n_i$.
The larger the number of cycles, the more complete the reconstruction of the tree will be, leading to more accurate statistics of the nodes and improving the quality of the final choice.
If the number of cycles is unbounded, MCTS is eventually equivalent to a brute force search and the policy converges to the optimal strategy~\cite{mcts_survey}.

\subsubsection*{The MCTS cycle.}

The four stages of an MCTS cycle are \textit{ (i) selection}, \textit{(ii) expansion}, \textit{(iii) rollout}, and \textit{(iv) backpropagation}.
\begin{itemize}
    \item[{\em i}.] In the \textit{selection stage}, starting from the root, the algorithm travels the part of the tree stored in memory by selecting for each node its child that maximizes a specified selection policy.  
The most common choice is the UCT (Upper Confidence
Bounds applied to Trees) policy \cite{Kocsis_MCTS, MCTS_review}, which selects node $a^*$ among the children of node $i$, $A(i)$, such that
\begin{equation}\label{eq:UCB1}
    a^* = \argmax{a\in A(i)} \left\{\frac{w_a}{n_a} + C \sqrt{\frac{2\ln n_i}{n_a}}\right\},
\end{equation}
where $n_i$ is the number of times the agent has visited the parent node $i$. The first term in \eqref{eq:UCB1} rewards the exploitation of well-scoring paths, while the second one encourages the exploration of less-known branches. The constant $C$ balances the two terms. The selection stage runs until a node with a child not yet added to the stored tree is selected.

\item[ {\em ii}.] In the \textit{expansion stage}, one unexpanded child of the node reached in the selection stage is expanded, i.e., added to the tree in memory with a score and a visit count of zero, and the algorithm travels to the newly expanded node.

\item[{\em iii}.] In the \textit{rollout stage}, random moves are played from the newly expanded node, without expanding them, until a leaf node of the original tree is reached. 
The reward associated with that leaf node is evaluated in this stage. 

\item[{\em iv}.] In the \textit{backpropagation stage}, the scores and visits are updated with the result from the rollout. The algorithm travels back to the root node increasing the visit counts of the nodes along the path by one and appropriately adding the reward obtained in the rollout to their cumulative scores.
\end{itemize}

The first two stages define the tree policy. They can be readily modified by changing the strategy of selection used in \textit{(i)}, and the way or the number of children that are expanded in \textit{(ii)} \cite{coquelin2007banditalgorithmstreesearch, Schadd_2008_MCTS_SP, chen_mctsQA}. 
The remaining two stages, where the leaf nodes are sampled and the expected rewards of different branches are updated, define the default policy of the algorithm. 
This policy can be modified by changing how the rollouts are carried out, such as by making the playouts follow some task-specific heuristics \cite{Winands_2011_modifiedplayouts, Schadd_2008_MCTS_SP, Cazenave2007_UCT_paralelization}. 
Overall, MCTS is independent of the specific problem it is facing, and can often be successfully used with very little knowledge of the corresponding action space. 
In cases where the standard version fails, the loose definition of its policies makes the algorithm highly flexible when developing problem-specific modifications \cite{mcts_survey, MCTS_review}. 

\subsection{MCTS for quantum optimization}
\label{sec:MCTS_for_QAOA}
Using MCTS for QAOA parameter optimization requires formulating the task as a sequential decision problem. 
A straightforward way to achieve this is to build a decision tree of depth $2P$ and branching factor $b$. At each depth $i\in [1,2P]$ the algorithm must choose a value for the parameter $\theta_i$, where $\bm{\theta}=\{\gamma_1, \beta_1, ..., \gamma_P, \beta_P\}$, out of $b$ discrete options, $\theta_i\in \{\theta_i^{(1)}, ..., \theta_i^{(b)}\}$, and the range of the parameters depends on the spectrum and the symmetries of the Hamiltonians generating the parameterized unitary operators.  
As we consider Hamiltonians with an integer spectrum, the variational parameters belong to the compact interval $[0, 2\pi)$.
Moreover, since the cost function is real, it satisfies $F_P(-\bm{\gamma}, -\bm{\beta})=F_P(\bm{\gamma}, \bm{\beta})$, 
meaning that, if the different $\theta_i^{(k)}$ are evenly spaced, for every path $\bm{\theta}$ there is an equivalent one where $\theta_i^{(k)}\rightarrow 2\pi -\theta_i^{(k)}$. 
Hence, we can restrict the domain of the first parameter to $[0,\pi]$, effectively halving the size of the tree to a total of $\left\lceil \frac{1}{2} b^{2P} \right\rceil$ leaf nodes, without changing the discretization step.

This approach transforms a continuous optimization problem into a discrete one; however, we do not expect it to affect QAOA performances significantly. 
Indeed, the algorithm's efficiency is limited by the complexity of the energy landscape and the structure of the decision tree, not by the impossibility of exactly pinpointing the target minimum because it does not fall on the discrete grid.
Yet, pathological cases might exist where extremely thin minima~\cite{Wauters_PRA2020,stechly2024} and wide flat regions require more clever discretization schemes than the regular ones.

Alternative formulations of the decision tree can also be devised, where the leaf nodes correspond to coefficients of a discrete basis for continuous functions with a compact domain. 
These can parameterize both smooth annealing schedules~\cite{chen_mctsQA} or QAOA strategies~\cite{Zhou_PRX2020}. 
Modulo minor differences depending on the specific parameterization of the schedule, the tree size will grow exponentially in the number of variational parameters. 
This is not a limitation for MCTS, as, in many cases, it is known to handle exponentially growing domains exceptionally well~\cite{MCTS_review}.

As we rely on the variational principle for the optimization, the reward collected at each leaf node must be related to the energy associated with the corresponding choice of parameters $F_P(\bm{\gamma},\bm{\beta})$. 
In this work, we chose a nonlinear mapping $r(\bm{\theta})=e^{-\nu F_P(\bm{\theta})}$ for the rewards, with $\nu$ chosen such that the reward starts rising when energies at or below the typical ranges are considered, and the algorithm can properly gauge the goodness of the parameters in the context of the problem at hand \footnote{To ensure the theoretical bounds on the regret of the playing agent \cite{UCB_proofs}, the UCT selection policy with $C=\sqrt{2}$ requires the rewards to lie in [0,1], although this is not strictly necessary for MCTS to perform well, and should be studied on a case-by-case basis \cite{Schadd_2008_MCTS_SP}.}.

In MCTS language, optimizing the QAOA parameters is a single-player game, meaning that no opposing agent is trying to reach a different goal.
This has some consequences for the optimal strategy choice: in a two-player game, a meaningful strategy is based on the assumption that the opponent is trying to win as well, and high-scoring paths based on counterproductive choices of the other agent are usually not worth exploring \cite{mcts_survey}.
This is not true in a single-player game, where the goal is maximizing the score considering all eligible strategies. 
In view of this, we implement some changes to MCTS proposed in Ref.~\cite{Bjornsson_2009_CadiaSP}. 
We first modify the criterion with which MCTS decides the final move to play, by choosing the child that leads to the path of the random simulation that obtained the highest reward. 
Second, we expand the highest-scoring path of each turn all the way to the leaf node in between turns, thus ensuring that MCTS memorizes such good paths in their entirety. 
The search still follows the UCT selection policy in~\eqref{eq:UCB1} for the selection stage, which can under or overestimate the potential of exploration of certain regions of the search domain.
But with these changes, we have the certainty that the final parameters MCTS chooses will be at least as good as those in the best memorized sequence. 
In the following, we will denote this modified variant as MCTS-SP (single-player). 

In the following sections, however, we will still consider the standard version of MCTS. 
First, it is interesting to see how it compares to the single-player variant and, second, we expect the single-player modifications to no longer be beneficial in some particularly relevant cases, such as when considering a noisy cost landscape (see Appendix~\ref{app:noise}). 
In this noisy situation, the task loses its deterministic nature, and the memorized parameter sequences will likely contain inaccurate expected rewards, making the standard final choice criteria more appropriate, and rendering path memorization irrelevant.

\subsection{Iterative Search-Space Restricted MCTS}\label{sec:SSR}
MCTS has been proven to work well for optimizing annealing schedules \cite{chen_mctsQA}, although gradient-descent algorithms perform slightly better in idealized settings (noiseless and smooth cost functions), thanks in part to the continuous parameter spaces~\cite{Wauters_2022}\footnote{We expect the real potential of MCTS as an optimizer to lie in robust handling of noisy or complex cost-landscapes where gradient-descent optimizers are prone to get stuck in local minima.}.
However, in our study, we found that standard MCTS is unsuited for QAOA as soon as $P>1$.
The modified single-player version obtains improved results but quickly fails to learn good strategies too, even in shallow circuits, as we will discuss later on in Sec.~\ref{sec:3sat}.
Therefore, MCTS needs to be adapted further to perform well in QAOA parameter optimization, leveraging on the structure of its energy landscape.

\begin{figure}
    \centering
    \includegraphics[width=0.6\linewidth]{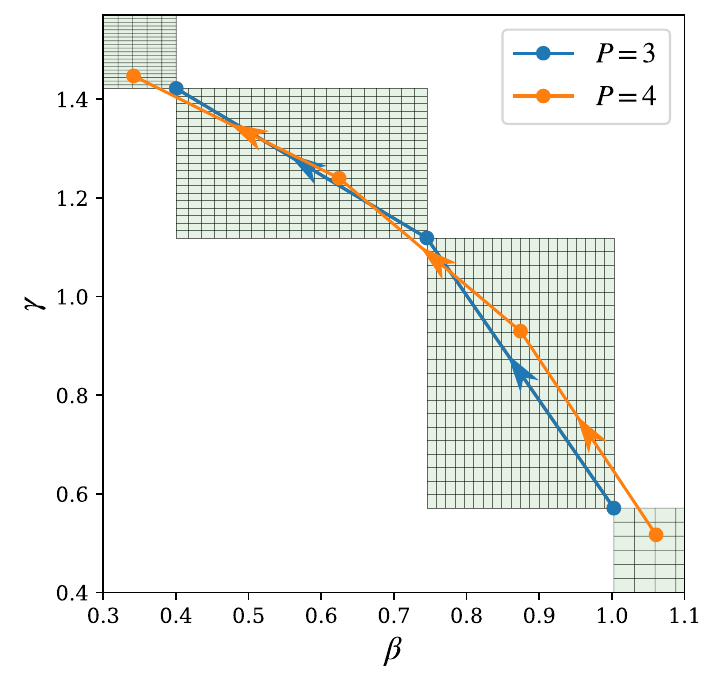}
    \caption{Example of a search space restriction step (with $\delta_4=0$ and $b=21$) when the depth is increased from $P=3$ to $P=4$. Each point in the plot represents a pair of parameters ($\gamma_i, \beta_i$) and the arrows indicate the direction of increasing $i$ for the parameter pairs. The SSR-MCTS algorithm only searches for the $P=4$ optimal parameters (orange) within the grid points of the highlighted green rectangles, which are generated from the $P=3$ optima (blue). In this example with $\delta_4=0$, all of the space outside of the green rectangles is not seen by SSR-MCTS.}
    \label{fig:SSR_example}
\end{figure}

In this section, we describe how to modify the standard MCTS algorithm to successfully tackle QAOA.
The core idea is not to modify the MCTS algorithm {\em per se}, but to progressively modify the space of eligible strategies. 
We call this method iterative search-space restriction (SSR). 
It consists of constraining the action space of the decision tree nodes as the circuit depth $P$ increases, according to the knowledge of previous solutions for $P-1$.
Indeed, QAOA optima often display regular structures and patterns in the optimal parameters~\cite{mbeng2019quantum,Zhou_PRX2020,Wauters_PRR2020, Akshay_PRA2021}, look ahead to Fig.~\ref{fig:3SAT_results}(b) for an example. 
This property stems from the presence of optimal discretized quantum annealing schedules, which are typically robust with respect to small parameter changes in the Hamiltonian, system size, and number of time steps.
These regular patterns appear as a monotonic increase (decrease) of the $\gamma_i$ ($\beta_i$) with the discrete-time index $i$, approaching a continuous function as $P\to \infty$.
Hence, when increasing the number of layers, the new optimal parameters typically resemble a finer interpolation of the optimal schedule for the previous circuit depth. 

For problem types/instances that exhibit these properties, we can implement an SSR-MCTS step in the following way: suppose that, for a depth $P$, the optimal parameters $\bm{\theta}^*=\{\gamma_1^*, \beta_1^*, ..., \gamma_P^*, \beta_P^*\}$ are known. 
We then let MCTS search for the optimal $P+1$ parameters $\bm{\tilde{\theta}}^*=\{\tilde{\gamma}_1^*, \tilde{\beta}^*_1, ..., \tilde{\gamma}^*_{P+1}, \tilde{\beta}^*_{P+1}\}$ in a mesh of $b$ discrete values per parameter within
\begin{align}\label{eq:searchspace_params}
    \tilde{\gamma}^*_i &\in [\gamma_{i-1}^*(1-\delta_{P+1}), \gamma_i^*(1+\delta_{P+1})], && \nonumber \\
    \tilde{\beta}^*_i &\in [\beta^*_{i-1}(1-\delta_{P+1}), \beta^*_i(1+\delta_{P+1})],
\end{align}
where $i\in[0,P+1]$. 
The values for $\beta^*_0$, $\gamma^*_0$, $\beta^*_{P+1}$ and $\gamma^*_{P+1}$ can generally be heuristically chosen values that limit the range of the new parameters, and $\delta_{P+1}$ is a \textit{softening} constant that specifies how much MCTS is allowed to search beyond the bounds of the restricted regions of parameter space for depth $P+1$. 
An example of a $P=3$ to $P=4$ search-space restricted search, with $\delta_{4}=0$, is depicted in Fig.~\ref{fig:SSR_example}. 

This method can be employed to look for the optimal QAOA angles one layer deeper when we know the optima for the previous depth, and its iterative application starting from $P=1$ allows for a systematic restriction of the action space by gradually increasing the number of parameters, without any prior knowledge of the optimal schedule. 
In this way, the parameters suggested by SSR-MCTS at depth $P$ are used to determine how the $P+1$ search is restricted.
The different $\delta_{P}$ for each step can be adjusted to relax the restriction and help the algorithm correct its potential inaccuracies in previous circuit depths \footnote{To make the method fully adaptive, one might introduce automatic ways to control the softening constants during the restriction procedure, which within the present implementation have to be set manually.}. 
In this iterative case, the boundaries of the first and last parameters ($i=0$ and $i=P+1$) do not need to be specified and can generally be unrestricted, i.e., set by the bounds of the original $P=1$ action space \footnote{In cases where their optimal values vary slowly with $P$ and high circuit depths are considered, the restricted regions of the first and last parameters would shrink much slower relative to the others, potentially causing them to lose the required resolution to select a value close to the optimal one. This could be easily circumvented by increasing (decreasing) the first (last) boundary values for the edge parameters to lie at most a fixed distance away from the previous optima, e.g., $\gamma^*_{P+1}$ would become $\gamma^*_{P+1}-\varepsilon(\gamma^*_{P+1}-\gamma^*_{P})$, for some $\varepsilon<1$.}.
This procedure is independent of the particular Hamiltonian instance $H_C$ one is facing: as the number of QAOA layers increases, it progressively generates a restricted action space tailored to the problem at hand, learning the structure of the parameter space from previous iterations.

As we will see in subsequent sections, SSR-MCTS makes the energy landscape seen by the MCTS agent easier to navigate,
as regions of the parameter space that are uninteresting or even hindering the optimization task are gradually discarded.
The complexity reduction of both the decision tree and the associated energy landscape is particularly evident when $P$ is large.
Indeed, the regularity regions become smaller as $P$ increases, while the branching factor remains unchanged, making the discretization of the parameters finer at each step.
Pictorially, it can be seen as zooming in the region around the target minimum with greater resolution at each iteration.
For this reason, and despite this progressive simplification of the decision tree, SSR does not affect its exponential scaling with $P$; in fact, after the first run, the restriction generally breaks inversion symmetry, effectively doubling the number of leaf nodes, with our formulation of the problem.

To keep the overall computational cost under control, we will consider MCTS with a number of function evaluations growing linearly with the circuit depth, leading to an overall quadratic scaling for iterative SSR-MCTS starting from an unrestricted $P=1$ search.

\section{Results - 3-SAT}\label{sec:3sat}

\subsection{The 3-Satisfiability problem}
As a first benchmark for the MCTS-guided quantum optimization, we consider instances of boolean 3-Satisfiability (3-SAT) problems.
3-SAT is one of the most well-known $\mathsf{NP}$-complete problems \cite{garey1979computers} and a standard benchmark for quantum optimization algorithms \cite{Battaglia_PRE2007,chen_mctsQA}. 
Given a set of $n$ classical binary variables $x_i\in \{0,1\}$,
it consists in finding whether a particular assignment $\bm{x}^*$ exists such that a boolean expression consisting of a logical conjunction of $m$ clauses 
\begin{equation}\label{eq:3sat}
    \varphi(\bm{x}) = \bigwedge_{a=1}^{m} C_a(\bm{x}) 
\end{equation}
is satisfied. 
In 3-SAT each clause is a disjunction of three literals $C_a = l_{a_1}\lor l_{a_2} \lor l_{a_3}$, with $l_{a_j}$ being $x_{i}$ or its negation $\neg x_{i}$, where $i=a_j$ is the index of the $j$th bit entering clause $C_a$.
The hardness of the satisfiability problem depends on the density of clauses $\alpha\equiv m/n$. It is in the vicinity of $\alpha_c\sim 4.3$, where 3-SAT has a well-known phase transition from satisfiable to unsatisfiable instances \cite{Mezard_Science2002}, that a backbone of strongly constrained variables emerges and one finds the hardest problem instances, with which existing 3-SAT solvers
struggle the most \cite{Weixiong2001Max3SAT}.

A given instance $\varphi$ can be encoded into a quantum Hamiltonian as 
\begin{equation}\label{eq:H-3sat}
    H_C = \sum_{a=1}^m\frac{1}{8}(\mathds{1}+g_{a_1}\sigma^z_{a_1})(\mathds{1}+g_{a_2}\sigma^z_{a_2})(\mathds{1}+g_{a_3}\sigma^z_{a_3}),
\end{equation}
where the sum runs over all clause indices, $\sigma^z_{a_j}$ represents the Pauli operator corresponding to the $j$th bit in clause $a$, and with $g_{a_j}$ taking the value $+1$ if the corresponding variable enters the problem statement \eqref{eq:3sat} as the negation $\neg x_{a_j}$, or $-1$ if it enters as $x_{a_j}$ instead. 
Equation \eqref{eq:H-3sat} generates long-range interactions with up to three-body terms, which require careful embedding to be encoded in present-day quantum hardware~\cite{Bastidas2020FullyprogrammableUQ}. 
Typically, they can be mapped with polynomial cost onto effective Ising Hamiltonians with equivalent low-lying states~\cite{Lucas2014}.

The expectation value of the cost Hamiltonian Eq.~\eqref{eq:H-3sat} gives the average number of unsatisfied clauses in the state under consideration.
Finding the ground state of said Hamiltonian is thus equivalent to solving the $\mathsf{NP}$-hard optimization version of the corresponding 3-SAT problem, also known as MAX-3-SAT~\cite{garey1979computers}, where the (harder) task is finding the bitstring that violates the minimum number of clauses. Clearly, this also solves 3-SAT, as the problem is satisfiable if and only if the ground state energy is zero. The phase transition of MAX-3-SAT happens for similar clause densities to $\alpha_c$, although it affects the hardness of the problem differently \footnote{The hardness of MAX-3-SAT also depends on the clause density $\alpha$, and one expects a phase transition to occur for some $\alpha\leq\alpha_c$, as MAX-3-SAT subsumes the complexity of 3-SAT. As $\alpha$ increases past $\alpha_c$ and the problem instances become more and more over-constrained, the hardness of MAX-3-SAT, unlike that of 3-SAT, never decreases. Such high clause densities make it easy to find that problem instances are unsatisfiable, but very hard to find their optimal bitstrings. We have gone from an \textit{easy-hard-easy} type of transition for 3-SAT to an \textit{easy-hard} one for MAX-3-SAT~\cite{Weixiong2001Max3SAT}}. 
Even though we are technically solving MAX-3-SAT, in the following we will refer to the problem instances as “3-SAT instances”, as it is customary for the sake of convenience.

In this work, we work with a set of 3-SAT instances with $\alpha=3$, each with a unique bit-string satisfying Eq.~\eqref{eq:3sat}. 
Despite this density of clauses being lower than the threshold for the complexity phase transition in the thermodynamic limit~\cite{Mezard_Science2002},
having a non-degenerate ground state makes the optimization hardness comparable to random problem instances in the critical regions~\cite{Znidaric_3SATlowalpha, chen_mctsQA}.
From a quantum annealing perspective, this is reflected in the quantum Hamiltonian having an energy gap that decreases exponentially with the qubit number, making annealing unfeasible for large systems.
3-SAT is known to be a formidable problem for QAOA, requiring large circuit depths to reach the ground state with high fidelity  ($P\gtrsim 45$ for 0.95 fidelity on 6 qubits~\cite{Akshay_QAOAreachability}).
However, as long as the projection over the ground state is large enough, an approximate optimization is sufficient to sample it with a finite probability.
As deep variational circuits are unfeasible for current hardware, we will limit the number of layers to $P \le 10$.
As a mixer Hamiltonian for QAOA we use a transverse field 
\begin{equation}
    H_M=\sum_i \sigma^x_i \ ,
\end{equation}
which we adopt throughout this paper. 
The code was developed using the open-source QuTip library \cite{qutip1,qutip2} and is available in \cite{my_github_page}.

\begin{figure}
    \centering
    \includegraphics[width=\linewidth]{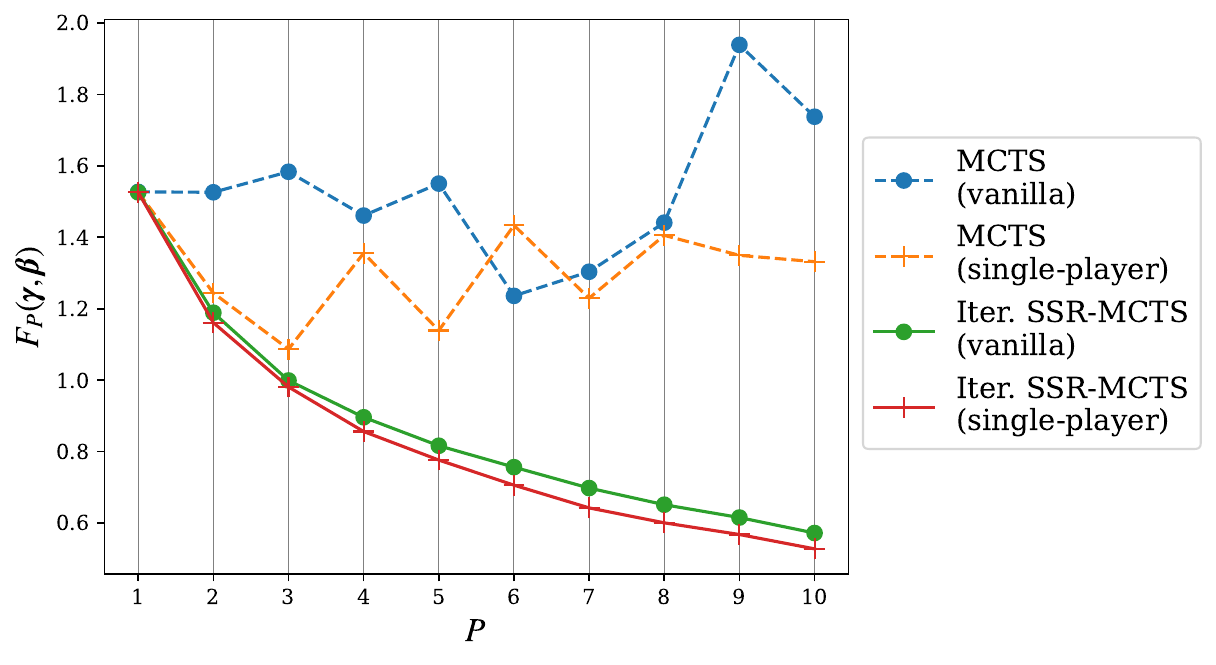}
    \caption{Comparison of the performance of different versions of MCTS on a $n=7$ example 3-SAT instance. Energies of the final states are shown as a function of circuit depth for vanilla MCTS (blue), MCTS with single-player modifications (orange), iterative SSR-MCTS (green), and iterative SSR-MCTS with single-player modifications (red). The number of MCTS cycles increases linearly with $P$.}
    \label{fig:mcts_comparisons}
\end{figure}

\subsection{\texorpdfstring{Unrestricted MCTS vs.\@ SSR-MCTS}{Unrestricted MCTS vs. SSR-MCTS}}
\label{sec:birdSSRcomparison}

Let us now test MCTS's ability to select good QAOA strategies for 3-SAT problems.
Based on the discussion presented in Sec.~\ref{sec:mcts}, we have four optimizer variants: vanilla MCTS, MCTS-SP, iterative SSR-MCTS, and iterative SSR-MCTS-SP.
In Fig.~\ref{fig:mcts_comparisons}, we report their performance, measured with the variational energy at the end of the optimization, as a function of the number of QAOA layers. 
The data refer to an example 3-SAT instance with $n=7$ qubits.
The most striking feature is that vanilla MCTS fails spectacularly as soon as $P>1$. The single-player variant mildly improves the performance, but again fails already for a modest circuit depth $P=4$.
On the other hand, iterative SSR completely changes the algorithm's effectiveness, making the ground state approximation more and more accurate as $P$ increases, and the single-player modifications now further improve the overall performance of iterative SSR-MCTS.
This behavior appears consistently across all tested instances and problem sizes.

\begin{figure*}
    \centering
    \includegraphics[width=\linewidth]{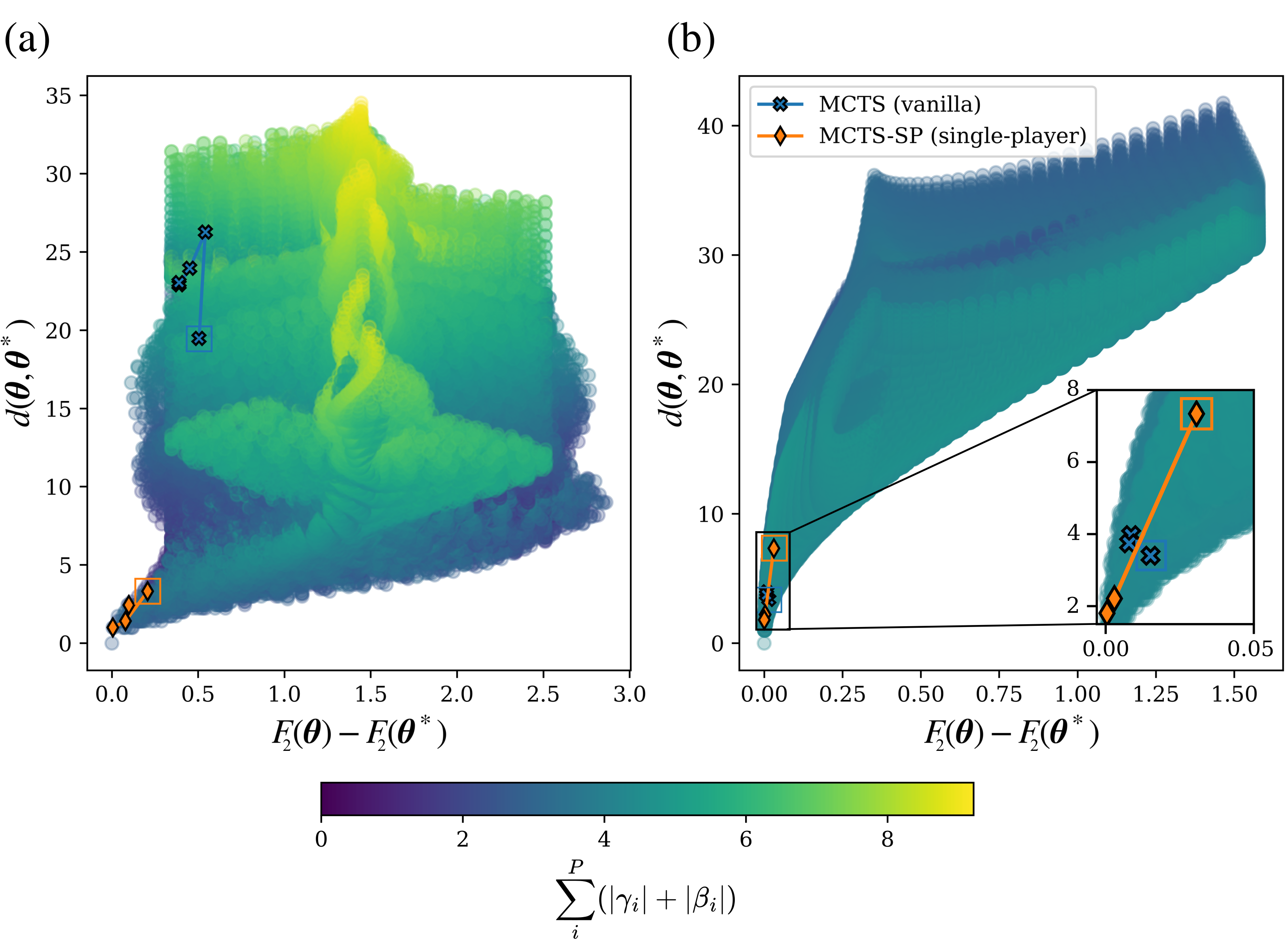}
    \caption{Distribution of all leaf nodes in the tree representing all possible combinations of parameters MCTS can choose for a circuit depth of $P=2$ and a branching factor $b=30$, in an example 3-SAT instance. (a) Leaf nodes from the full parameter space (up to inversion symmetry), (b) leaf nodes within the restricted parameter space defined according to the $P=1$ optimal angles. 
    Each leaf node is represented according to its difference in energy and tree distance (defined in~\ref{sec:treedistdef}) from the optimal node $(F_2(\bm{\theta})-F_2(\bm{\theta}^*),d(\bm{\theta},\bm{\theta}^*))$, and the color of each leaf node represents the sum of the corresponding QAOA parameters (run time). The nodes corresponding to the best possible sets of parameters are the ones on (0,0), the bottom-left corners of the figures. For each of the two cases, we also show the five leaf nodes selected by both standard MCTS (blue crosses) and MCTS-SP (orange diamonds) after five runs, each with different numbers of MCTS cycles per turn. The symbols in the squares correspond to the results with the fewest number of cycles, and the lines connect the leaf nodes in order of increasing cycle numbers in the set $\{200, 600, 1000, 1800, 3500\}$.}
    \label{fig:leaf_distr}
\end{figure*}

The dramatic effect of SSR already at $P=2$ is due to its impact on the leaf-node structure. To appreciate it, in Fig.~\ref{fig:leaf_distr} we compare the energy distribution among the leaf nodes for a depth $P=2$ when MCTS has the unrestricted $[0,2\pi)^{2P}$ space to search in (a), with SSR generated from the $P=1$ optima in (b).
Each leaf node can be identified by a tuple composed by its energy difference to the optimal choice $F_P(\bm{\theta})-F_P(\bm{\theta}^*)$, the corresponding tree distance $d(\bm{\theta},\bm{\theta}^*)$ (a measure of the deviation from the path of the optimal leaf, defined in~\ref{sec:treedistdef}), and the effective run time $\tau = \sum_i(|\gamma_i|+|\beta_i|)$.
Figure~\ref{fig:leaf_distr} shows the relation between these quantities.

Comparing the two panels, the effect of SSR emerges strikingly, as it deeply modifies the correlation between energy and distance from the optimal strategy.
In panel (a) there are many leafs with high energy but a small distance from the absolute minimum, meaning many low-energy nodes can be effectively hidden among siblings with undesirable energies. 
This makes gauging the quality of different moves very challenging for MCTS. Furthermore, in the case of the standard variant, even when the algorithm eventually finds a path leading to some optimal leaf node, its contribution to the exploration of the tree in future cycles and to the final chosen move is almost inconsequential: the optimal leaf node was most likely sampled very late in the search to modify the nodes' average scores (and subsequent search paths) significantly. 
Indeed, this feature is exactly one of the main factors that make MCTS struggle in single-agent contexts~\cite{MCTS_review}, suggesting that its performance might benefit from tailored policy modifications \footnote{There are also cases in which the standard algorithm performs well in single-player domains, look at~\cite{Bjarnason_Fern_Tadepalli_2009, Kocsis_MCTS, MCTS_puzzles, MCTS_singleplayerworks}, but unrestricted QAOA parameter optimization, as a hard deterministic single-player game, happens to not be one of those games.}.

If we look at SSR in panel (b), instead, a large energy difference also corresponds to a node far from the optimum. 
This means that good paths down the tree are more concentrated in particular branches and can be reached via similar parameter choices in the early turns of the game, making MCTS able to discover and learn strategies to navigate the cost landscape adeptly.
The energy interval spanned by the possible QAOA schedules is also reduced, notice the difference between the scales on the $x$-axis. SSR gets rid of all terminal nodes corresponding to excessively large run times and highly excited states, thus focusing on the parameter region where fast-running and low-energy minima reside.
More details on the leaf node structure are given in Appendix~\ref{app:ssrtree}.

The difficulties of vanilla MCTS persist if one tries to improve its performance by increasing the number of cycles.
An example is provided by the set of blue crosses in both panels of Fig.~\ref{fig:leaf_distr}. 
Starting from 200 MCTS cycles per turn (boxed cross) we increase this number to $\{600, 1000, 1800, 3500 \}$. The links connect MCTS results with a growing number of cycles.
When MCTS has to search through the whole parameter landscape, panel (a), it typically gets stuck on leaf nodes far away from the optimal one.
On the contrary, it consistently finds well-scoring nodes near the optimal one for the SSR case.

Single-player modifications (orange diamonds) can partially cure this problem, or enhance the performance of SSR-MCTS.
In the problem represented in Fig.~\ref{fig:leaf_distr}, MCTS-SP approaches the optimal leaf node as the number of cycles is increased both with and without SSR. 
This modified version, however, relies on finding the optimal leaf nodes within the rollout stages, which we expect to become exponentially harder in the non-SSR case, as the number of leaf nodes and local minima increases with circuit depth. 
This makes this single-player modified version only usable in practice to find the minima for low-depth QAOA unless iterative SSR is employed to find regular parameters; indeed, its performance plummets with increasing depth, and it fails already for $P=4$ (see Fig.~\ref{fig:mcts_comparisons}) unless we allow the number of cycles to grow exponentially in $P$.

\subsection{3-SAT with Iterative SSR-MCTS}
\label{sec:3SATiterSSRresults}
Now that we had a glimpse into the importance of search-space restriction for transforming the QAOA decision tree into one that MCTS can efficiently find optimal strategies in, let us delve into more details of the iterative SSR-MCTS protocol we use. 
In each run, the algorithm performs 1000 initial cycles plus 800 more per turn, excluding the final one, while selecting the parameters. The last parameter to choose does not require further stochastic rollouts since the best move can be selected straightforwardly. 
Hence, even though the search space scales exponentially with $P$, the number evaluations of $F_P(\bm{\gamma}, \bm{\beta})$ only increases linearly with growing circuit depth, $n_{\rm fev} = 1000 + 800(2P-1)$, as two new turns are added at each subsequent $P$. 
See Appendix~\ref{app:cycles} for an analysis of the performance dependence on the number of training cycles.
We set $C=\sqrt{2}$ in the selection policy and $\nu=1/2$ for the exponential mapping between energy and reward discussed in \ref{sec:MCTS_for_QAOA}.
At each depth $P$, the range of the parameters is bounded according to the parameters proposed for the previous depth, and a branching factor of $b=30$ is used to build the tree within the restricted space. 
The boundaries are softened with the values $\bm{\delta} =
\{0, 0, 0.1, 0.05, 0.04, 0.03, 0.02, 0.01, 0.01\}$, where $\delta_P$ indicates how much the search is allowed to search outside of the regular regions at depth $P$, according to \eqref{eq:searchspace_params}.
We fix $\delta_1=0$ since we are already considering the entire space at the shallowest depth, and $\delta_2=0$ because we expect the algorithm to find the $P=1$ optimal parameters in every case. 
We set a nonlinear decrease of $\delta_P$ with the circuit depth when $P\ge 2$.
\begin{figure}
    \centering
    \includegraphics[width=\linewidth]{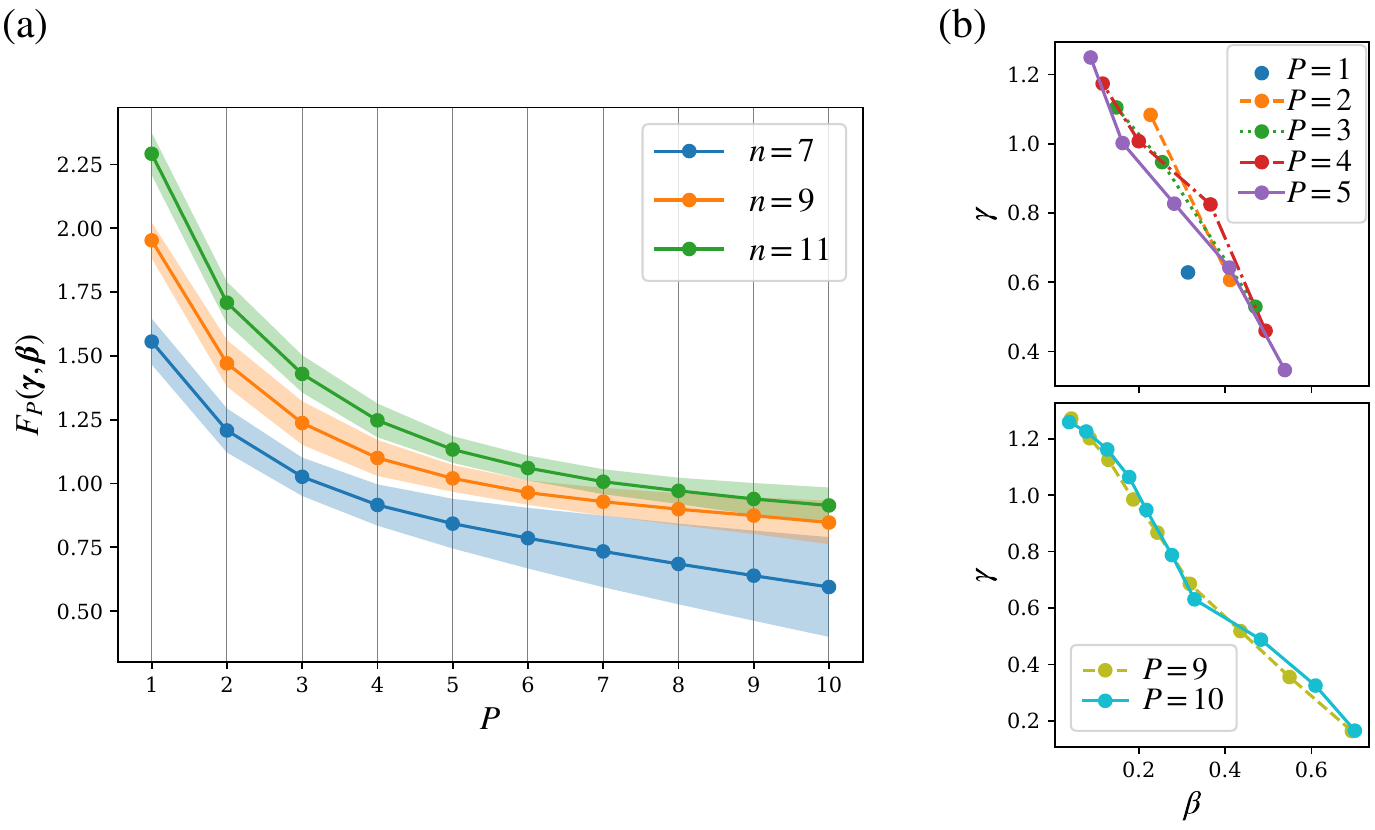}
    \caption{(a) Average energies obtained with the iterative SSR-MCTS algorithm for different problem sizes for 3-SAT as a function of the circuit depth. The energies for each $n$ are averaged over 15 problem instances, each with a clause to variable ratio $\alpha=3$ and one unique satisfying ground state. 
    The algorithm performs $1000+800(2P-1)$ cycles at depth $P$. (b) Example QAOA angles chosen by iterative SSR-MCTS to approximate the ground state of an example $n=9$ 3-SAT instance at different circuit depths. The parameters from $P=1$ to $P=5$ are shown on the top panel, and the $P=9$ to $P=10$ ones on the bottom panel. The visualization of the parameters is similar to that of Fig.~\ref{fig:SSR_example}, with the parameter indices increasing from bottom-right corners to top-left ones.
    }
    \label{fig:3SAT_results}
\end{figure}

Figure~\ref{fig:3SAT_results}(a) reports the results for QAOA with iterative SSR-MCTS for different 3-SAT instances of varying problem sizes $n=7,\ 9,\ 11$ and $m=3n$ clauses. 
Each data point is the average over 15 problem instances, while the shaded areas indicate their standard deviation.
The algorithm performs reliably as $n$ increases, as shown by the monotonic decrease of the cost function $F_P(\bm{\gamma}, \bm{\beta})$ as a function of the depth $P$.
The energy remains high, centered around the lowest excitation, because of the known reachability deficit for QAOA applied to satisfiability problems~\cite{Akshay_QAOAreachability}.

The effect of SSR on the QAOA schedule can be appreciated from Fig.~\ref{fig:3SAT_results}(b), where we show an example set of parameters chosen by iterative SSR-MCTS for different depths on a 3-SAT instance with $n=7$.
Each dataset is a sequence $(\beta_i,\gamma_i)$ with the first pair corresponding to the point closer to the lower-right corner of the plot.
As $P$ increases, the optimal parameters clearly approach a regular behavior, linked to optimal digitized-quantum-annealing schedules and known to be associated with optimal minima strategies robust with respect to small parameter and size variations~\cite{mbeng2019quantum,Zhou_PRX2020,Wauters_PRR2020}.
While iterative SSR is the analog of other iterative optimization methods of QAOA parameters~\cite{mbeng2019quantum,Zhou_PRX2020} in the context of MCTS, it is remarkable that it completely changes standard MCTS performance already at $P=2$. While the regularity at large circuit depth is expected, its appearance is noteworthy when the search space is only moderately restricted.

\subsection{Noisy rewards}
\label{sec:noisyrewardsmc}
A key aspect of any optimization strategy is its capacity to handle noisy energy landscapes. 
This is particularly relevant in practical implementations of Variational Quantum Algorithms (VQA), where expectation values are derived from measurements with stochastic outcomes. 
Additional qubit errors in the quantum circuit further complicate the search for optimal parameters. 
In this work, we focus solely on measurement noise and defer a more comprehensive analysis to future research.
As monitoring uncertainty can be mitigated by increasing the number of shots, in the following, we explore the robustness of SSR-MCTS under varying levels of measurement noise.

\begin{figure}
    \centering
    \includegraphics[width=0.86\linewidth]{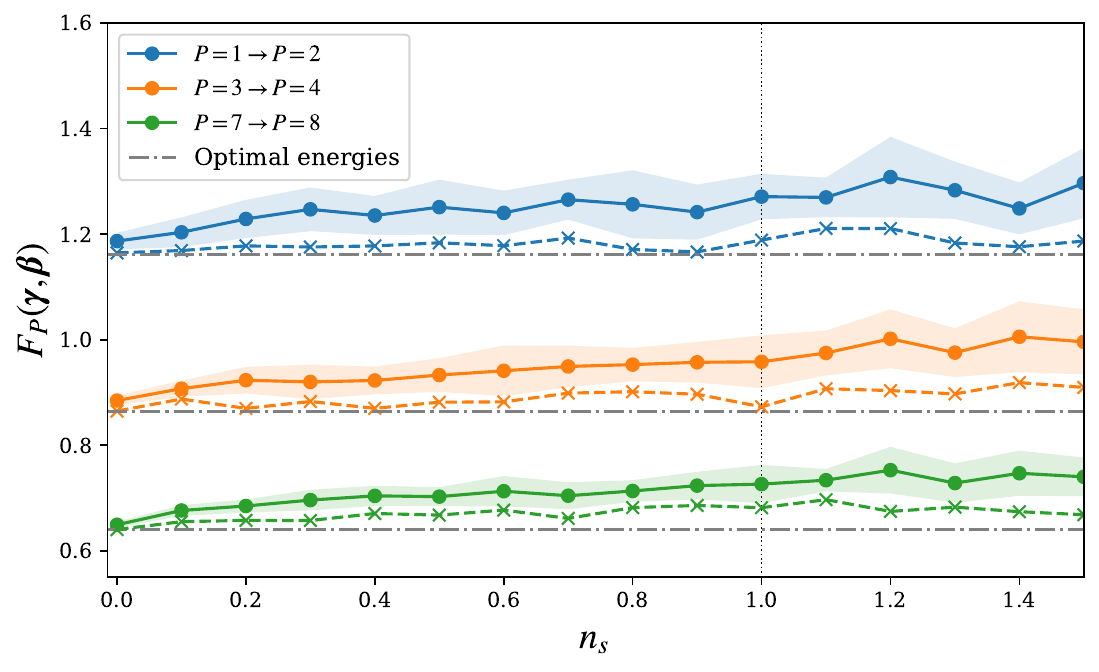}
    \caption{Energies given by the parameters chosen by SSR-MCTS for a hard instance of the $n=7$ 3-SAT problem with a noisy cost function. Results from three different steps are shown, $P=1$ to $P=2$ (blue), $P=3$ to $P=4$ (orange), $P=7$ to $P=8$ (green), each with varying noise strength $n_s$. 15 SSR-MCTS runs were performed for each noise configuration. The mean (circles with solid lines), the standard deviation (shaded areas), and the best game (crosses with dashed lines) are shown for each. The horizontal dot-dashed lines represent the optimal energy obtained at each step for $n_s=0$ and the vertical line marks $n_s = \Delta$.}
    \label{fig:noise}
\end{figure}
We model the measurement uncertainty by introducing a random Gaussian variable to the rewards during the rollout phase of each MCTS cycle. 
Consequently, the algorithm must select variational parameters based on noisy data accumulated in the statistics of the decision tree. 
Specifically, we use random variables $\mathcal{N}(0,n_s)$ with zero mean and a standard deviation determined by the noise level $n_s$. 
This addition modifies the reward function as $F_P(\bm{\theta}) \rightarrow F_P(\bm{\theta}) + \mathcal{N}(0,n_s)$. 
For evaluation purposes, the final assessment does not include this random noise.
MCTS generally handles noisy rewards well though its effectiveness can vary based on factors like noise intensity, problem structure, and algorithm implementation~\cite{mcts_survey}.
The relevant energy scale for comparing noise strength is the typical energy gap $\Delta$ between eigenvalues. 
As long as $n_s \ll \Delta$, MCTS should effectively distinguish between leaf nodes leading to states close to different eigenstates, as the node scores are already subjected to uncertainty due to stochastic sampling.  
However, if the noise strength approaches the gap $n_s \sim \Delta$, it can, in principle, significantly affect the estimation of optimal actions, making it difficult to differentiate between strategies.
In our 3-SAT formulation (see Eq.~\eqref{eq:H-3sat}), the energy units correspond to the number of violated clauses, with the minimum relevant energy gap being $\Delta = 1$.

To test the protocol's robustness against noise, we focus on the hardest 3-SAT instance (smallest energy gap in a QA setting) with $n=7$ and repeat 15 SSR-MCTS runs, without the single-player modifications (see Appendix~\ref{app:noise} for a comparison with other variants and COBYLA~\cite{cobyla}), for 3 different jumps on $P$ between 1 and 8, and $n_s \in [0,1.5]$.
Remarkably, MCTS has no issues finding the optimal parameters as it is able to average out the noise over different cycles even for large noise strengths $n_s \gtrsim \Delta$, as we show in Fig.~\ref{fig:noise}. 
As the noise strength is increased, the average performance gets mildly worse (full circles with solid lines) but it is often able to find very good parameters in one of the runs. Indeed, the best results out of the 15 repetitions (crosses with dashed lines) drift away even more slowly than the average from the noise-free QAOA energy, as $n_s$ increases.
This remarkable robustness makes iterative SSR-MCTS well-suited for applications in noisy quantum devices, as it does not rely on gradient estimations and the uncertainty on the reward estimate can be easily taken into account when designing how the leaf nodes are sampled.

\section{Results - Unweighted MaxCut}\label{sec:maxcut}

\begin{figure}
    \centering\includegraphics[width=0.86\linewidth]{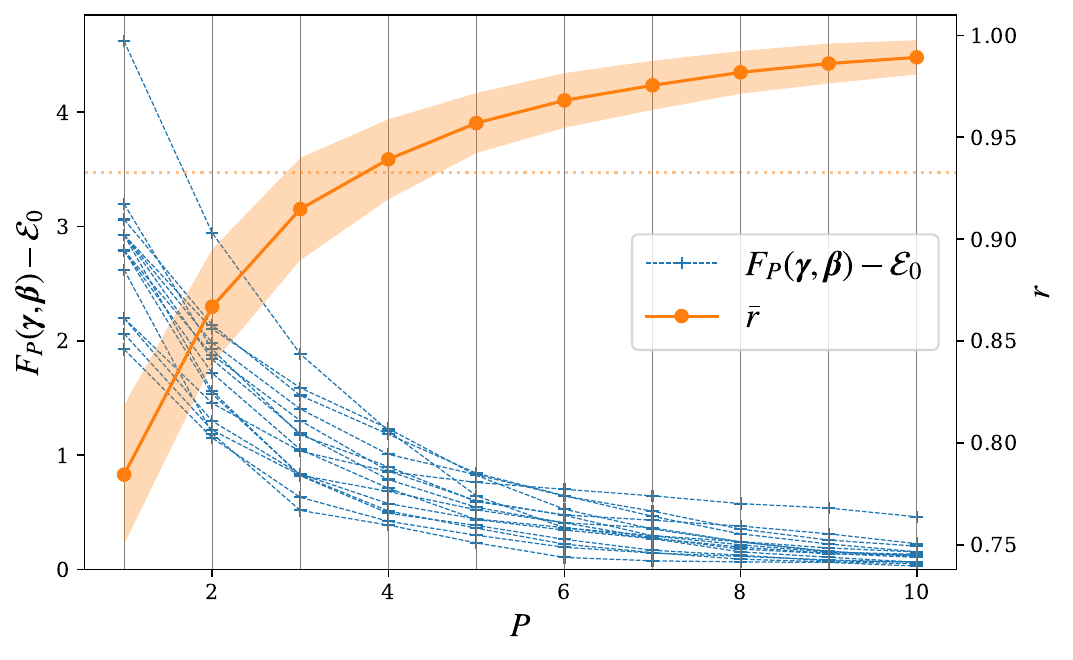}
    \caption{Energy differences from ground states (dashed blue curves) and the average approximation ratio $\bar{r}$ (orange curve) obtained with iterative SSR-MCTS vs the circuit depth $P$ for MaxCut problems on unweighted 3-regular graphs with $n=10$ vertices. The data was obtained using the 19 possible non-isomorphic connected 3-regular graphs with 10 vertices. The dotted horizontal line represents the performance guarantee of the  Goemans-Williamson algorithm for 3-regular graphs, $r^*\approx 0.9326$ \cite{Goemans1995, Halperin2004}.}
    \label{fig:MaxCut_results}
\end{figure}

To confirm that our results are not unique to 3-SAT, we repeat the performance analysis on unweighted MaxCut instances.
MaxCut is another prototypical $\mathsf{NP}$-hard optimization problem, extensively studied in QAOA contexts \cite{Farhi_arXiv2014, Zhou_PRX2020, Wurtz_PRA2021, Herrman2021, Wang_PRA2018}. 
It consists in classifying the vertices $V$ of a graph $G(V,E)$ into two sets such that the number of edges in $E$ joining vertices between the different sets is maximal. 
The problem Hamiltonian for MaxCut can be expressed as an antiferromagnetic Ising model on $G$,
\begin{equation}
    H_C = \sum_{<i,j>\in E}\frac{1}{2}(\mathds{1}+\sigma^z_i\sigma^z_j),
\end{equation}
with each ferromagnetic edge carrying an energetic penalty of 1 unit. 
The size of the cut for a configuration $\ket{\bm{\sigma}}$ with energy $\mathcal{E}=\braket{\bm{\sigma}|H_C|\bm{\sigma}}$ on a graph with $N_E$ edges will be given in this representation by $C(\mathcal{E})=N_E-\mathcal{E}$. The ground state energy $\mathcal{E}_0$ thus gives the maximum cut $C_{\text{max}}\equiv C(\mathcal{E}_0)$. 
A common figure of merit for MaxCut is the approximation ratio $r$, which for an energy $\mathcal{E}$ is defined as
\begin{equation}
\label{eq:approxratiomc}
    r(\mathcal{E}) = \frac{C(\mathcal{E})}{C_{\text{max}}},
\end{equation}
such that $r=1$ if the optimal cut is found.
In Fig.~\ref{fig:MaxCut_results} we show the results of iterative SSR-MCTS for 19 MaxCut problem instances on $n=10$ unweighted 3-regular graphs. 
We plot both the energy for each individual instance (blue crosses) and the average approximation ratio over all instances (orange circles).
The latter is compared to the performance guarantee of the Goemans-Williamson (GW) algorithm~\cite{Goemans1995, Halperin2004}, which is a standard classical reference for MaxCut problems.
The convergence to the ground state is smooth with increasing values of $P$, similar to what we observed for 3-SAT problems.

The accuracy of iterative SSR-MCTS is qualitatively similar to the results obtained with warm-start gradient descent approaches~\cite{Zhou_PRX2020}.
At $P=4$, both optimization strategies of QAOA schedules beat, on average, the GW algorithm.
At larger values of $P$, when $E\to 0$, the discretization of the parameter space marginally limits the performance of our method, preventing the algorithm from reaching the exact ground state ($r=1$) even if reachable by the variational QAOA ansatz. For instance, at $n=10$ and $P=10$, we find $1-r\simeq 0.01$, one order of magnitude larger than Ref.~\cite{Zhou_PRX2020}.
A larger branching factor in the decision tree would result in a finer resolution of the parameters, which mitigates this limitation at very small residual energies, at the cost of making the tree exploration more resource-expensive. Furthermore, improved search-space restriction schemes for the edge parameters, such as searching a limited distance from the previous optima as suggested in Sec.~\ref{sec:SSR} or making the discretization non-uniform (finer closer to the previous optima), could be another step in fully optimizing the performance of iterative SSR-MCTS for deep circuits.
In general, however, we can not expect MCTS to outperform gradient-based algorithms working with continuous parameters in ideal simulations where the cost-function is noiseless {\em and} the energy landscape has only a few smooth minima, such as the regular MaxCut instances tested here. 
In experiments, instead, it can be advantageous because of its resilience against unavoidable noise. Furthermore, with single-shot measurements,  sampling the ground state with high probability does not require perfect fidelity with the final variational state.

\section{Conclusions and outlook}\label{sec:conclusion}

In this paper, we analyzed the performance of MCTS as a gradient-free optimizer for QAOA variational parameters.
Despite its success in designing optimal schedules for quantum annealing~\cite{chen_mctsQA}, vanilla MCTS fails when faced with the complex energy landscape of QAOA.
This complexity emerges as fractal-like decision trees, characterized by the close proximity of low- and high-energy leaf nodes, where it is impossible to discriminate between good and bad branches in polynomial time reliably.
We overcame these difficulties by restricting the search space of QAOA parameters based on the knowledge of optimal parameters obtained for shallower circuits and leveraging the well-known presence of regular schedules in QAOA~\cite{mbeng2019quantum,Zhou_PRX2020,Wauters_PRR2020,Akshay_PRA2021, Streif_QST2020, Sack_TQA_init_QAOA}.
By testing it on 3-SAT and MaxCut instances, we showed that iterative SSR-MCTS provides a systematic way to optimize QAOA parameters within the MCTS framework.
Preliminary numerical results indicate its performance is comparable with other heuristic optimization methods while offering at the same time great resilience against noise, suggesting it effectively accommodates both the potentiality and limitations of NISQ applications.

Moreover, the flexibility of MCTS suggests further possible development.
First, further modifications to the tree policy can be envisioned to improve the performance and tailor it to specific VQA applications.
For instance, the algorithm would likely benefit from changes to the selection criterion itself \cite{coquelin2007banditalgorithmstreesearch}, such as using a different kind of policy entirely \cite{Chaslot_MCTSSP, cazenave2009nested} or modifying the UCT to favor the exploration of nodes with a large reward variance~\cite{Schadd_2008_MCTS_SP, Schadd2011thesis}.
MCTS can also be efficiently integrated with neural networks~\cite{alphazero_original,chen_mctsQA} to speed up the nodes' score evaluations and improve the transferability between different problem instances or sizes. 
This will likely make the algorithm more robust at the price of a greater computational cost in the training phase.
Finally, we expect MCTS to be paired well with many modifications of QAOA as well as other VQAs. 
Being problem agnostic and relying only on a well-defined reward (cost) function, which is a key ingredient of VQAs, iterative SSR-MCTS represents a reliable workhorse to add to the toolkit of hybrid quantum-classical algorithms.

Optimizing VQA parameters is generally a challenging task due to features such as barren plateaus~\cite{McClean_QAOA_barren_plateaus,Arrasmith_Quantum2021} and the proliferation of local minima~\cite{Lumia_PRXQuantum2022}, which hinder efficient exploration of the energy landscape.
Warm initialization of variational parameters has been shown to mitigate these difficulties~\cite{Egger2021warmstartingquantum,Tate2023warmstartedqaoa,Truger_2024}, and iterative SSR-MCTS essentially follows a similar philosophy. 
Despite this common ground, standard variational optimization and MCTS approach the problem from fundamentally different perspectives:
The former treats QAOA as a function optimization problem over a high-dimensional manifold, while the latter frames it as a sequential decision-making process.

This distinction motivates further studies to investigate how the structural properties of the decision tree in MCTS relate to the geometry of the energy landscape itself. 
In particular, could the observed “fractalization” of the decision tree be indicative of the roughness or flatness of the energy landscape~\cite{stechly2024}? 
Understanding such connections could offer deeper insights into the intrinsic difficulty of variational optimization, as well as reveal new ways to exploit decision-tree structures to navigate complex landscapes more effectively. 
Indeed, one might argue that the hardness of QAOA optimization is an inherent property of the algorithm rather than a consequence of the optimization method used. 
However, reinforcement learning approaches have been shown to uncover regular patterns in QAOA parameters that remain elusive to traditional warm-started optimizations~\cite{Wauters_PRR2020}. 
Given that MCTS can be viewed as a crude form of reinforcement learning, it is possible that it could outperform direct energy landscape minimization in certain cases where conventional optimizers struggle the most.

Furthermore, alternative strategies such as Bayesian optimization~\cite{frazier2018BO} and surrogate methods~\cite{forrester2008engineering,Kubler_quantum2020} have recently gained traction as powerful and robust tools for the classical loops of VQAs~\cite{Tibaldi_2023,Shaffer_PRA2023,Cheng_CommPhys2024}. A detailed comparison of their respective strengths and limitations, as well as an investigation of possible fruitful combinations of these methods, would be instrumental in developing a more comprehensive toolkit for hybrid quantum-classical algorithms, enabling the selection of the most effective optimization method for a given problem.

Another crucial aspect in the development of quantum algorithms for NISQ devices is their resilience against errors and decoherence. These, indeed, typically degrade the performance of VQAs when the circuit depth increases, even if the optimization loop is entirely done on classical machines~\cite{Pelofske_2023}.
For more realistic applications, it is therefore important to understand whether iterative SSR-MCTS can cope with hardware noise and still suggest meaningful strategies.

Addressing these issues will advance our understanding of the mutual enhancement between quantum computing and artificial intelligence, and foster the development of quantum technologies.

\section*{Data availability statement}
The data that support the findings of this study are openly available at~\cite{agirre_arabolaza_2024_13970332}.

\begin{acknowledgments}
We warmly thank M. Burrello and G. Giedke for useful discussions and advice. 
AA acknowledges funding by the Department of Education of the Basque Government through the project PIBA\_2023\_1\_0021 (TENINT), and that this work has been produced with the support of a 2023 Leonardo Grant for Researchers in Physics, BBVA Foundation. The BBVA Foundation is not responsible for the opinions, comments and contents included in the project and/or the results derived therefrom, which are the total and absolute responsibility of the authors.
MW has received funding from the European Union’s Horizon Europe research and innovation programme under grant agreement No 101080086 NeQST.
This project has been supported by the Provincia Autonoma di Trento and Q@TN, the joint lab between the University of Trento, FBK—Fondazione Bruno Kessler, INFN—National Institute for Nuclear Physics, and CNR—National Research Council.
Views and opinions expressed are however those of the author(s) only and do not necessarily reflect those of the European Union or the European Commission. Neither the European Union nor the granting authority can be held responsible for them.
This work was supported by the Dutch National Growth Fund (NGF), as part of the Quantum Delta NL programme.
\end{acknowledgments}

\appendix

\section{MCTS and gradient-descent: on possible hybrid optimizer variants}\label{app:globalopt}

MCTS is an extremely versatile algorithm. The very loose definition of its selection and playout policies make it very convenient to modify to suit the problem at hand or develop new alternative algorithms \cite{mcts_survey, MCTS_review}. In this appendix, we showcase two of the many promising avenues opened by this approach for developing MCTS-based optimizers to find optimal QAOA strategies. 
The core idea is to use MCTS to survey at a coarser resolution the QAOA energy landscape while a gradient-based local minimization, for instance the BFGS algorithm~\cite{BFGS_book}, is tasked with finding the local minima.
We look at two possible implementations.

\begin{figure}
    \centering
    \includegraphics[width=0.6\linewidth]{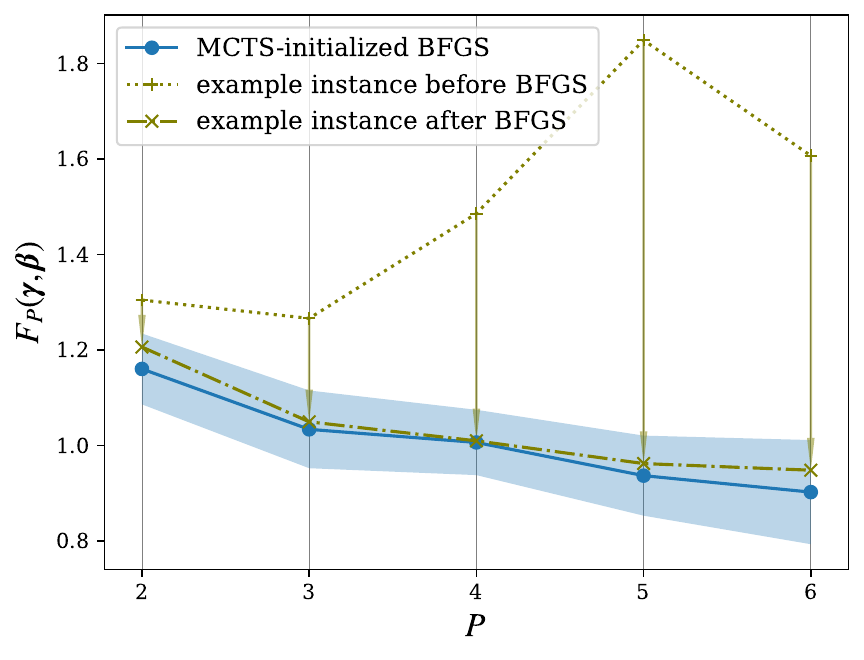}
    \caption{Average energies obtained with MCTS-initialized BFGS searches on 10 $n=7$ 3-SAT instances (blue). 10 games of standard (non-SSR) MCTS were run per instance, and the parameters suggested by MCTS were used as initial parameters of a BFGS search. Only the best result out of the ten games of each instance contributes to the average. For one example instance, the initial energies given by the MCTS search are shown (green ``$+$” markers), suggesting that parameters chosen by MCTS with large residual energies can still be on the slope to a minimum with much smaller energy (green ``$\times$” markers) in many cases.}
    \label{fig:mcts_initialized_bfgs}
\end{figure}
\begin{figure}
    \centering
    \includegraphics[width=0.6\linewidth]{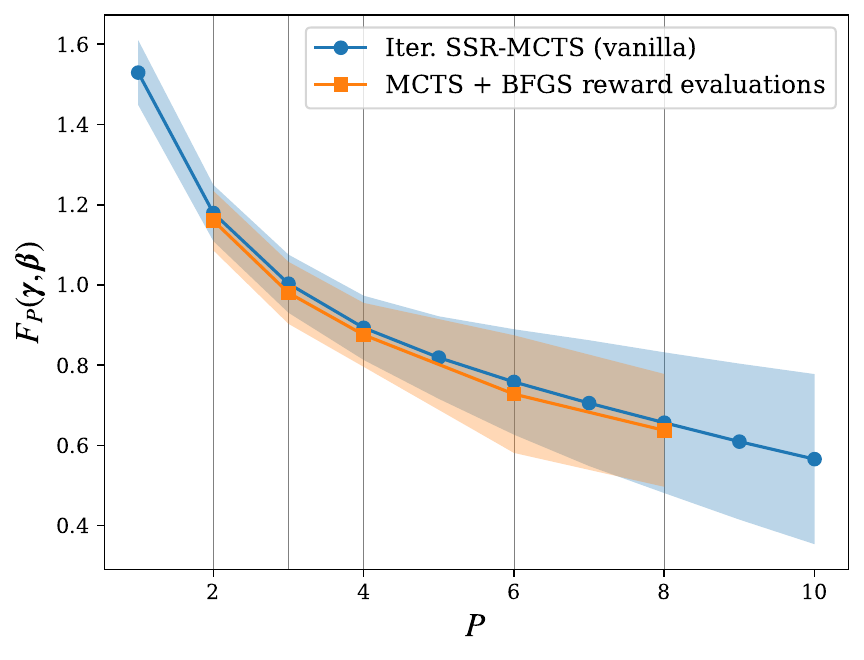}
    \caption{Energies of MCTS with BFGS reward evaluations (orange) and standard MCTS with iterative SSR (blue) averaged over 10 $n=7$ 3-SAT instances. For this BFGS variant, a gradient descent search was run at each leaf node reached during the rollout stages of MCTS. In this way, the algorithm sees a landscape made up of different flat plateaus with the values of all the minima, which correspond to the regions from where the gradient points to that particular minimum. MCTS ran for just $100$ cycles per turn for each instance.}
    \label{fig:bh_and_SSR}
\end{figure}

In the first, as proposed in \cite{chen_mctsQA}, MCTS simply suggests initial parameters for a subsequent gradient-descent search, even when the schedule chosen by MCTS gives rise to a high energy. 
Indeed, providing ways to efficiently initialize local (gradient-based) optimization is very valuable in the context of the complex cost landscapes of QAOA, as it is well-established that randomly initialized QAOA tends to converge to undesirable local minima \cite{Sack_TQA_init_QAOA}.
We present a benchmark of this approach in Fig.~\ref{fig:mcts_initialized_bfgs}, which shows the results of MCTS-initialized gradient-descent searches using the standard BFGS \cite{BFGS_book} algorithm on 10 different instances of $n=7$ 3-SAT. For each instance, we consider the best result out of 10 runs of MCTS-initialized QAOA. 
The MCTS-initialization improves the results considerably from standard MCTS, showing that the parameters suggested by unmodified MCTS can, in fact, sometimes be on the slope leading to a good minimum, and not necessarily stuck in a local minimum like the high energy seems to suggest.
This approach relies exactly on what makes MCTS fail in the first place, i.e. the ``fractalization” of the leaf nodes. While it is hard for MCTS to identify an optimal strategy, it could potentially identify something that is close enough in the parameter space to easily find it with a local minimization with continuous parameters initialized at that point. 
The combined resources for the MCTS initialization and the subsequent BFGS gradient descent, however, make this approach's computational cost scaling unfavorable for large parameter spaces.

The second alternative hybrid algorithm can be devised by modifying the rollout stage of MCTS. 
In this version of MCTS, the reward of a leaf node corresponding to a set of QAOA angles, is instead extracted from the energy given by the parameters obtained via a BFGS search starting from the parameters of the original leaf node. 
Consequently, this version of MCTS sees the cost landscape as a collection of “flat” surfaces, each with a cost corresponding to its particular minimum, and extending to the whole basin of attraction of said minimum. This is somewhat reminiscent of the basin-hopping algorithm for multi-variable optimization \cite{Wales1997}, with the difference that the hopping dynamics occur naturally with MCTS.

We show the results obtained with this second algorithm in Fig.~\ref{fig:bh_and_SSR}, comparing it to the iterative SSR-MCTS for 10 $n=7$ problem instances. 
We find MCTS does not struggle within such landscapes for the depths under consideration, at least with the problem size we are considering. 
The good performance obtained by MCTS in these types of basin-based landscapes can be attributed to the fact that MCTS tackles the exploration of the energy landscape in a different way than most local optimizers. Indeed, the latter start from a random point in the parameter space and then update all variational parameters (or a batch of them) by looking at the surrounding landscape. 
MCTS, on the other hand, sees the sequence $(\bm{\gamma}, \bm{\beta})$ as a particular strategy on a decision tree, where at each layer the next move is chosen to maximize the probable future reward and is less prone to get stuck in flat regions, as it samples other areas of the landscape and focuses on the more promising regions. This variant also especially benefits from the single-player modifications outlined in Sec.~\ref{sec:MCTS_for_QAOA}. Concretely, sampling the lowest energy plateaus just once already implies that the optimal set of parameters will be found by the algorithm.

However, running MCTS in this manner greatly increases the number of required function evaluations for the same number of MCTS cycles, as each rollout requires a gradient-descent run. Nevertheless, the simplified landscape also seems to reduce the number of cycles required for MCTS to choose good strategies, although this does not make up for the vast increase in function evaluations given by the gradient-descent runs. With 100 MCTS cycles per turn, this algorithm is able to match the performance of iterative SSR-MCTS for $n=7$ 3-SAT, as seen in Fig.~\ref{fig:bh_and_SSR}. 
This algorithm is, in a sense, similar in concept to the one used in \cite{Yao_PRX2021}, where MCTS is used to propose Hamiltonian sequences for creating generalized QAOA ans\"atze, and the goodness of a sequence is gauged by the energy of the final state obtained with the parameters outputted by a gradient-based method for the given sequence; except, in our case, the sequence is fixed (we use the standard QAOA one) and we employ this method which utilizes gradient-descent runs during the rollout stages to optimize the parameters for that fixed sequence.

Both of these hybrid proposed methods have the advantage that the QAOA parameters they suggest are no longer discretized and that they are not bound by any problem-dependent heuristic considerations. However, the MCTS algorithm loses its main advantages, such as its gradient-free nature and remarkable resistance against noise.

\section{Why standard MCTS struggles: structure of the decision tree with and without SSR} \label{app:ssrtree}

\begin{figure*}
    \centering
    \includegraphics[width=\linewidth]{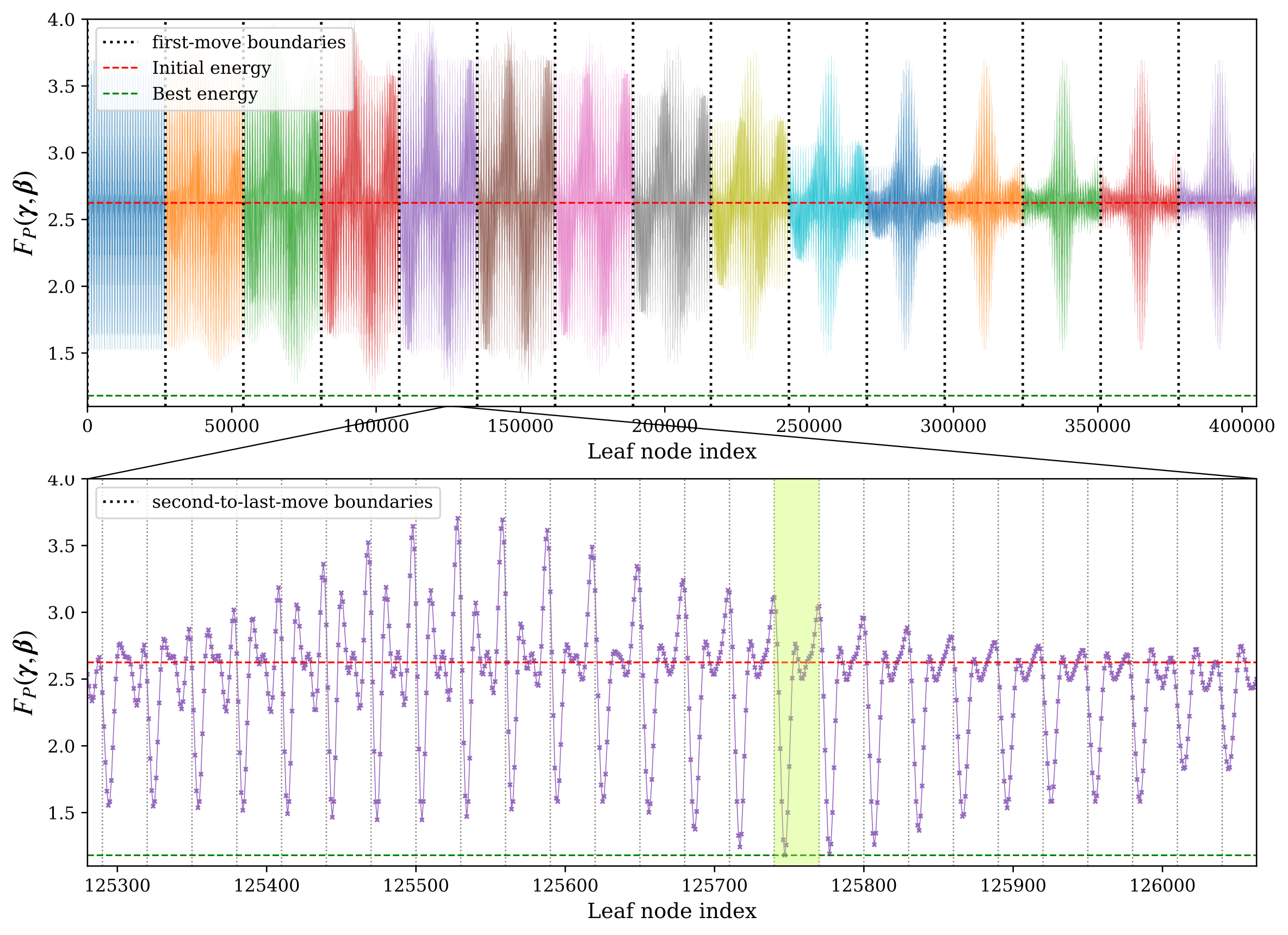}
    \caption{(Top) All of the leaf nodes and their energy for $P=2$ with an unrestricted decision tree ($b=30$) for an example 3-SAT instance of $n=7$ variables. The nodes are arranged as explained in Sec.~\ref{sec:MCTS_for_QAOA}. The choice of the first parameter constrains MCTS to one of the regions surrounded by the first-move boundaries (dotted vertical lines). The energy of the initial ($\ket{+}^{n=7}$) state (red dashed line) and the energy of the best leaf (green dashed line) are also shown. (Bottom) The leafs close to the optimal leaf node and their corresponding energies. Here the second-to-last move boundaries are shown (dotted vertical lines), and the region with leaf nodes with the optimal variational parameters up to the second-to-last one is highlighted in yellow. If the algorithm plays the move that takes it to the highlighted leaf nodes, it will have the option of choosing the optimal set of parameters by making the correct last choice for the remaining parameter within the highlighted region. The fact that the best-scoring leaf nodes are to the left of each second-to-last move sector (corresponding to small $\beta_2$), agrees with what we would expect from such a structure that benefits from an SSR, which sets an upper bound on $\beta_2$.}
    \label{fig:allleafs}
\end{figure*}
\begin{figure*}
    \centering
    \includegraphics[width= \linewidth]{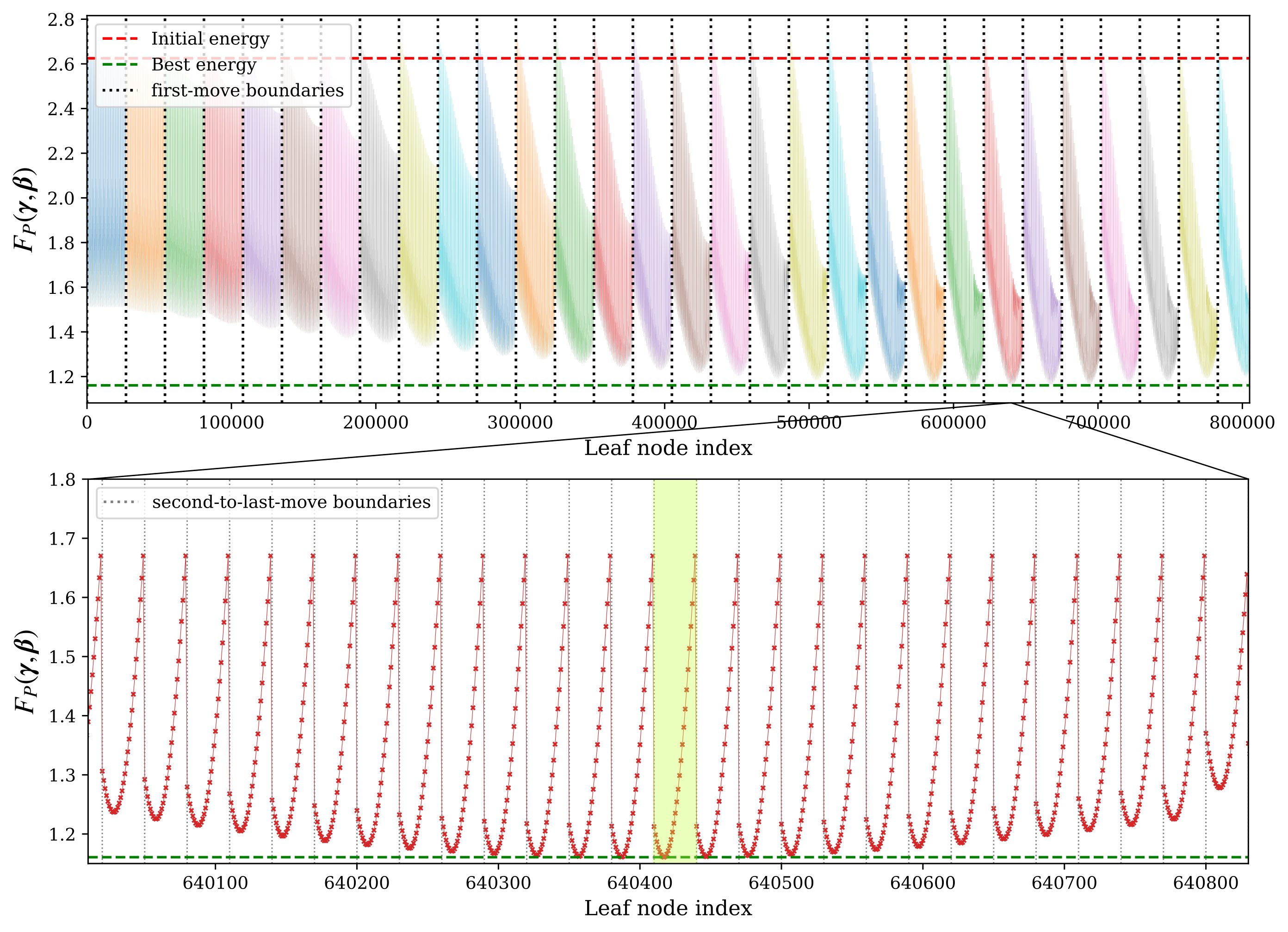}
    \caption{(Top) Energy difference from the optimal leaf node for all leaf nodes of the SSR $P=2$ decision tree ($b=30$) for an example 3-SAT instance with $n=7$ variables. The regions with fixed $\gamma_1$ are enclosed by the first-move boundaries (vertical dotted lines), and the energy of the optimal reference leaf (green dashed line) and the energy of the initial superposition state (red dashed line) are also shown. (Bottom) Energy of the terminal nodes around the optimal leaf. The second-to-last-move boundaries (dotted vertical lines) represent the regions where every parameter except the last one has been fixed, and it shows the energies obtained with each of the $b=30$ values of the remaining parameter. The optimal second-to-last move, which contains the optimal leaf node is highlighted in yellow.}
    \label{fig:ssr_all_leaf_nodes}
\end{figure*}

In this appendix, we extend the analysis of the $P=2$ cost landscape that was discussed in Sec.~\ref{sec:birdSSRcomparison}. 
As we found out, SSR is a requirement for standard MCTS to perform well in the task of QAOA parameter optimization. 
We expect this vast difference in performance between the two versions to manifest as clearly distinct decision tree structures.
We now look at the MCTS trees directly for the case of the example $n=7$ 3-SAT instance that we considered in Fig.~\ref{fig:leaf_distr} in order to study how the action spaces themselves differ in the unrestricted and SSR cases.

Fig.~\ref{fig:allleafs} shows the energy difference from the optimal set of parameters for all of the $4.05\times 10^5$ available leaf nodes (possible combinations of the variational parameters) for the unrestricted case. In Fig.~\ref{fig:ssr_all_leaf_nodes} we represent the same feature for the $\delta_1=0$ SSR, with the restriction implemented according to the previous depth's optimal parameters.
Notice that there are twice as many ($8.1\times 10^5$) possible outcomes in the SSR case since inversion symmetry can no longer be straightforwardly exploited. 
For both cases, we show the energies of all leaf nodes and represent the corresponding first-move boundaries, which enclose the regions that MCTS must choose between in the first turn, i.e., sets of leaf nodes sharing the same value of the first parameter ($\gamma_1$ in our case). 
In addition, we also present a zoomed-in version of both figures, showing the leafs close to the terminal node corresponding to the best approximation of the ground state within the variational manifold.
In these zoomed-in figures, we also show the second-to-last move boundaries, the regions where all but the last parameter ($\beta_2$ in this case) have been fixed \footnote{The different second-to-last boundaries in Fig.~\ref{fig:allleafs} (bottom) and Fig.~\ref{fig:ssr_all_leaf_nodes} (bottom), correspond to cutaway sections with constant $\gamma_2$ of the remaining 2D cost landscape when all but the last two parameters have been specified. This gives rise to an almost self-similar-looking structure, which slowly varies between different sectors.}.

Visually, the two cases are vastly different. The unrestricted case shows a very complex and unpredictable distribution of leaf nodes, where good strategies are likely hard to find. The best nodes are hidden in regions where very badly scoring ones also exist, and the averaging of the UCT strategy makes standard MCTS blind to these features in favor of paths that lead to mediocre leafs more consistently.
The improvements due to SSR can be argued qualitatively already from the structure of the energies in Fig.~\ref{fig:allleafs}. There is one key observation: to restrict this $P=2$ search space, we make use of the optimal parameters $(\tilde{\gamma}_1^*,\tilde{\beta}_1^*)$ for the same instance at $P=1$. As such, $\tilde{\gamma}_1^*$ sets an upper bound on $\gamma_1$, which means that the entire right part of the leafs (those with $\gamma_1>\tilde{\gamma}^*_1$) in Fig.~\ref{fig:allleafs} will be discarded. Indeed, we see that these turn out to all be undesirable leaf nodes. Furthermore, $\tilde{\beta}_1^*$ sets a lower bound for $\beta_1$, which in turn, means that the left part within each first-move sector is filtered out, and, again, we see that it is the parts with high $\beta_1$ that house the best minima. This qualitative argument can be extended all the way to the choice of the last parameter. 

In the SSR case of Fig.~\ref{fig:ssr_all_leaf_nodes}, in contrast, visual inspection suffices to see that the leaf node structure becomes much easier for MCTS to navigate: the first moves' quality is straightforward to estimate, with the behavior of the leafs in each sector being consistent and more regular, compared to Fig.~\ref{fig:allleafs}. In particular, there are no hidden good minima among leaf nodes scoring worse on average. 
Notice also that the parameter restriction already preselects a region of the variational manifold where the energy is always smaller than the initial one, marked by the dashed red line.
As we saw, standard MCTS can learn to very efficiently navigate such spaces.

As a closing remark, it is important to note that a high difference in the index of two leaf nodes of these figures need not necessarily indicate that the parameters of the two leaf nodes are far away from each other in Euclidean distance in the space of the variational parameters. For this purpose, representing the leaf nodes as we did in Fig.~\ref{fig:leaf_distr} is more appropriate.
Below we give more details on the definition of the tree distance employed in Sec.~\ref{sec:birdSSRcomparison}.

\subsection{Definition of tree distance} \label{sec:treedistdef}
We end this appendix with the formal definition of the tree distance used to generate Fig.~\ref{fig:leaf_distr}. For a depth $i$, there are $b$ possible values to choose for the corresponding variational parameter $\theta_i$,
\begin{equation}
    \theta_i \in \{\theta_i^{(1)}, ..., \theta_i^{(b)}\}.
\end{equation}
A leaf node is defined by a set of $2P$ such choices, the indices of which we represent with $c_i\in[1,b]$,
\begin{equation}
    \bm{\theta}=\{\theta_1^{(c_1)}, \theta_2^{(c_2)}, ..., \theta_{2P}^{(c_{2P})}\}.
\end{equation}
We define the tree distance $d$ between two leaf nodes given by configurations $\bm{\theta}$, $\bm{\theta}'$ as the euclidean distance on the indices $c_i, c'_i$ of their defining parameters,
\begin{equation}
    d(\bm{\theta}, \bm{\theta}')^2=\sum_i^{2P}(c_i-c'_i)^2.
\end{equation}
This is related to the (non-periodic \footnote{The periodic version of the tree distance can also be considered, although slightly harder to implement in the SSR case. For our analysis, both the periodic and non-periodic versions gave qualitatively similar results.}) distance of the two points in $2P$ dimensional parameter space.

\section{MCTS variants in the midst of noise}
\label{app:noise}

\begin{figure}
    \centering
    \includegraphics[width=\linewidth]{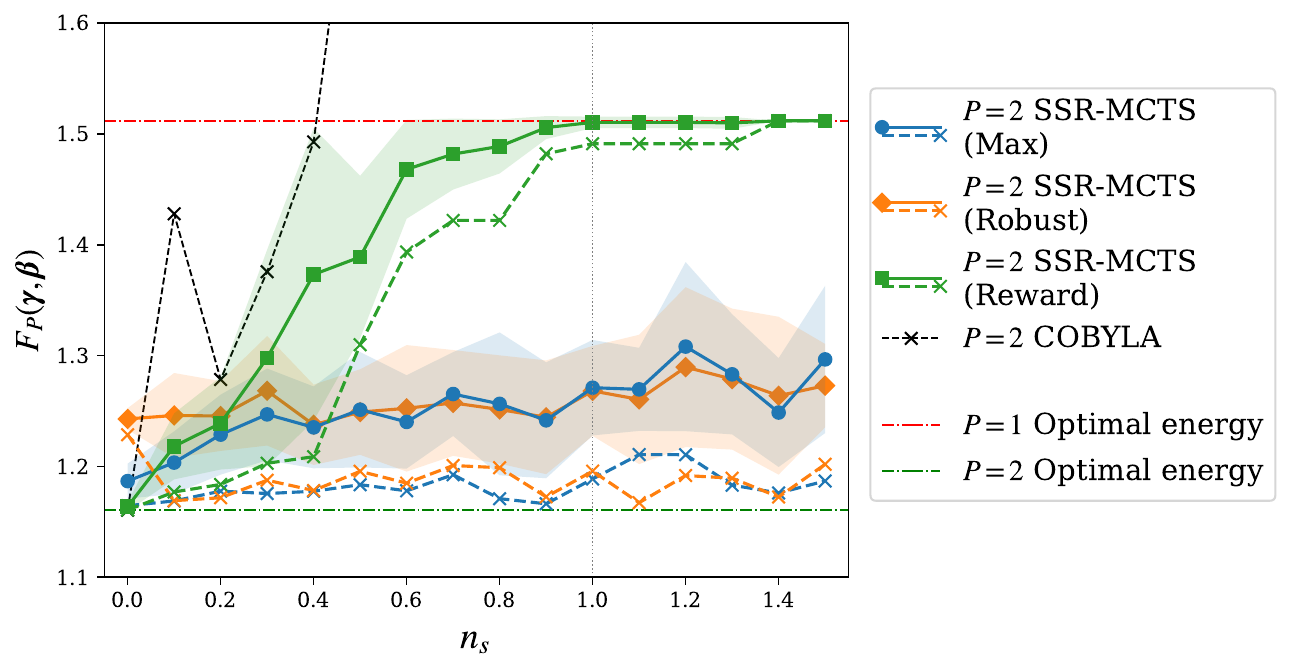}
    \caption{Comparison of the performance of different MCTS variants with varying noise strength. Average energy obtained with the standard \textit{max child} (blue circles), \textit{robust child} (orange diamonds) and single-player variant (green squares), where the child leading to the highest obtained \textit{reward} is chosen. The crosses of each color represent the best game for each noise setting for the corresponding variant. We also show the lowest energy obtained by COBYLA (black crosses) among 20 runs for each noise strength (shown only up to 0.4). The SSR-MCTS results are averaged over 15 games for each noise configuration. The results correspond to an example 3-SAT instance with $n=7$. The horizontal red and green lines indicate the optimal energy given by the $P=1$ and $P=2$ optimal angles, respectively. }
    \label{fig:variant_noise_comparison}
\end{figure}

In the main text, we utilized standard SSR-MCTS (i.e., without the single-player modifications) to optimize noisy cost functions (Sec.~\ref{sec:noisyrewardsmc}). In this appendix, we briefly look at how other versions of our algorithm fare within noisy landscapes.
Apart from the standard MCTS algorithm (the one which makes the final decision by choosing the child with maximum $w_i/n_i$, also known as \textit{max child}), we also test the standard algorithm with another widespread final move selection criterion known as \textit{robust child} \cite{Chaslot2008robust, mcts_survey, Chaslot2008progressive}, which chooses the child with the highest visit count $n_i$, as well as the single-player variant introduced in Sec.~\ref{sec:MCTS_for_QAOA}.
We also compare the different MCTS approaches with COBYLA~\cite{cobyla}, a gradient-free optimizer widely used in VQAs and as a standard baseline comparison to other optimizers~\cite{Lavrijsen2020, Cheng_CommPhys2024, Egger_PRR2023}, despite not being particularly optimized for noisy landscapes \cite{Cheng_CommPhys2024, Pellow-Jarman2024}. 
We do not provide an explicit comparison with BFGS, a standard gradient-based benchmark algorithm for noiseless optimization, as we found it unable to converge even at the smallest noise setting that we considered.

Fig.~\ref{fig:variant_noise_comparison} shows how each of the three tested variants of MCTS and COBYLA perform in finding the optimal parameters for search-space restricted $P=2$ QAOA on a single instance of 3-SAT with $n=7$ qubits and increasing noise strength. 
For the MCTS variants, we report both the average performance with the associated variance (continuous lines with solid marks and shaded areas) and the best result obtained out of 15 repetitions (dashed lines with crosses). For COBYLA we only report the best result out of 20 runs (dotted black lines with crosses).
In the noiseless case, the single-player variant performs best out of the MCTS-based variants as expected, as it is able to reliably store paths that lead to well-scoring sets of parameters. It also matches the quality of the COBYLA optimization, even though the latter explores a continuous energy landscape.
However, their performance quickly falls off as the noise strength is increased past $0.2\Delta$: the memorized paths of MCTS-SP are no longer a reliable criterion for choosing moves, and COBYLA becomes extremely unstable. 
Indeed, The single-player variant of MCTS is designed for deterministic action spaces, a quality that is lost as soon as we add noise or errors into the reward evaluation. 
Interestingly, for strong noise, the single-player variant converges to the energy of the $P=1$ optimal parameters.
We do not report data for COBYLA beyond $n_s=0.4$ as some points fall far outside the shown y-axis range.

The other two variants perform similarly, with a decisive improvement over the single-player variant and COBYLA at strong noise. In the $n_s=0$ case, \textit{robust child} seems to consistently choose worse schedules than the other variants. 
In principle, \textit{robust child} does not directly take the scores of the nodes into consideration for the final move selection. This is a potential advantage in noisy environments as the nodes' scores are affected by noise directly, which might suggest it might be more stable than \textit{max child}. However, the scores also do affect how the nodes are chosen in the selection stage, which, in turn, influences the visit counts. 
The worse performance of robust-child in the noiseless case might also indicate that the algorithm could benefit, at least in this example case, from alternative energy-reward mappings or different $C$ parameter values, to increase the reward difference between different leaf nodes to make the search greedier and less horizontal, and visit counts across sibling-nodes more varied.

\section{Resource cost analysis}
\label{app:cycles}

\begin{figure*}
    \centering
    \includegraphics[width=\linewidth]{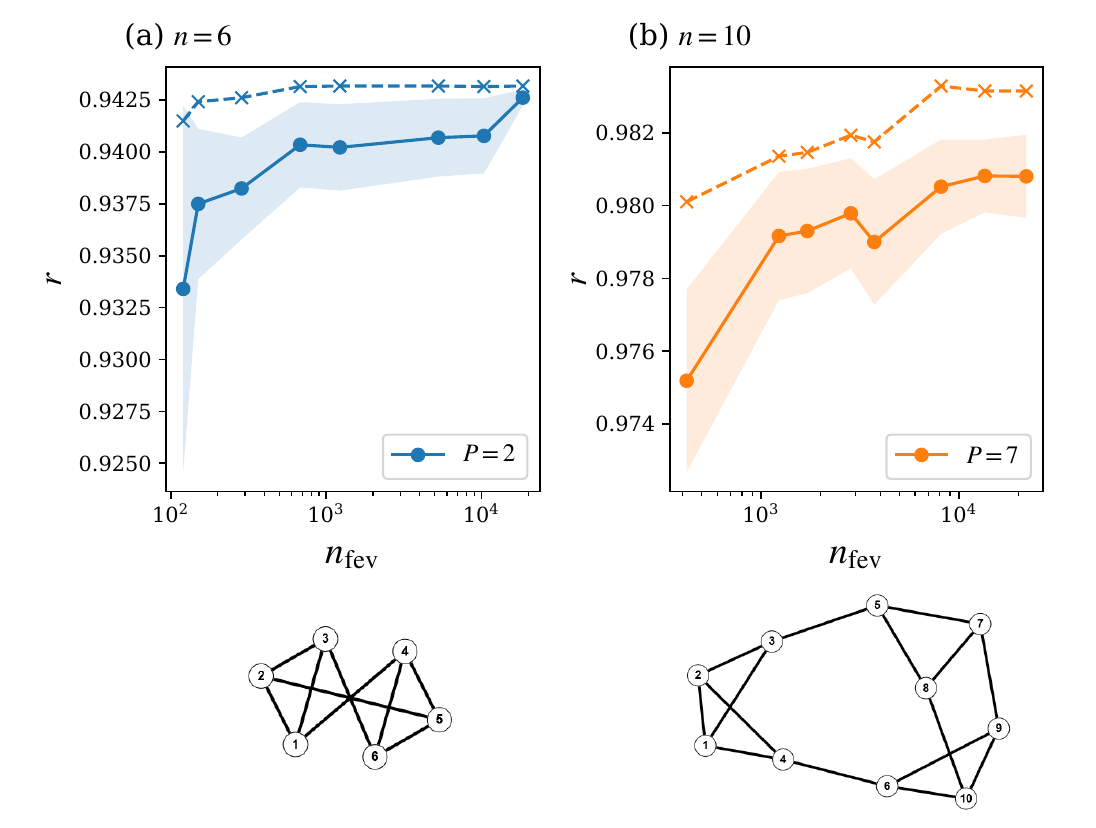}
    \caption{Average approximation ratio for MaxCut vs the total number of function evaluations during the optimization process for the MaxCut problem on two unweighted 3-regular graphs: one with $n=6$ vertices tackled with a QAOA ansatz of depth $P=2$ (a), and another with $n=10$ and $P=7$ (b). The data are averaged over 30 runs for each cycle number, the shaded regions show the standard deviation and the crosses indicate the best approximation ratio obtained out of the 30 games. The approximation ratios are calculated from the expectation values of $H_C$ with \eqref{eq:approxratiomc}. Below each of the two plots the corresponding unweighted 3-regular graph is shown. These were generated using \cite{graphonlinepage}.}
    \label{fig:rvscycles}
\end{figure*}

When designing optimization strategies for VQAs, the number of required queries to the quantum hardware is an important figure of merit for assessing their performance. 
This also limits the circuit depths that are realistically achievable, as a larger number of variational parameters also results in a larger required number of function evaluations.

As an \textit{anytime} algorithm \cite{mcts_survey}, the number of times the cost function is evaluated $n_{\rm fev}$ can be very easily controlled with MCTS. The number of calls matches the number of MCTS cycles performed during the game, times the number of shots needed for the desired accuracy, and they can be stopped at any point to make the algorithm suggest a strategy. 
The minimum number of cycles that we can run without leaving out any options are $2Pb$ (the depth of the tree times the branching factor), where the selected move at each turn will be the child node which gave the highest score after running one random simulation from each of the $b$ children. 
Evidently, the results from such MCTS runs are bound to be very inaccurate. We also know that, as the number of cycles grows to infinity, the algorithm eventually finds the optimal leaf node \cite{mcts_survey}.
In general, the training of machine-learning and AI methods is resource expensive and we expect MCTS to be no exception. In the main text, we used a heuristically chosen scaling $n_{\rm fev}=1000+800(2P-1)$. 
Here we elucidate the performance dependence on $n_{\text{fev}}$, which can be used to balance between the algorithm's cost and its accuracy.
Importantly, the $n_{\text{fev}}$ cycles do not need to be evenly distributed among the $2P$ choices that constitute the final strategy. Indeed, we find that spending more time at the beginning of the process is beneficial, as the first moves are critical for designing a successful schedule. Instead, optimizing the parameters ``deeper'' in the ansatz requires less exploration of the remaining part of the tree.

Figure~\ref{fig:rvscycles} shows the performance of iterative SSR-MCTS with varying number of total MCTS cycles and system sizes.
The results in Fig.~\ref{fig:rvscycles}(a) are obtained with $\delta=0$ on a graph with $n=6$ vertices (shown below the main plot), while the softening is $\delta=0.02$ in Fig.~\ref{fig:rvscycles}(b), where we run the algorithm with $P=7$ layers on a graph with $n=10$ (also shown below the plot).
The branching factor is set to $b=30$ in both cases. The sizes of the trees in the two cases thus amount to $8.1\times10^5$ and $2.2\times10^{10}$ leaf nodes, respectively, out of which the algorithm samples only $n_{\text{fev}}$ (not necessarily unique) nodes.
The two panels display an improving performance as $n_{\text{fev}}$ increases, both in the average and in the best-found approximation ratio. However, the range in which $r$ varies is rather limited, below $2\%$, and it would be significantly overshadowed by the presence of noise. Hence, a few hundred cycles are enough to reach good approximations of the target state, as the search-space restriction already limits the variational ansatz in a region close to the global minimum. 
In particular, notice that in Fig.~\ref{fig:rvscycles}(b) SSR-MCTS already reaches $r\ge 0.95$ for the minimum number of cycles considered $n_{\text{fev}}=420$.
Our results suggest that iterative SSR-MCTS is competitive against other state-of-the-art algorithms, such as the Bayesian Optimization in ref.~\cite{Tibaldi_2023}. The performance can be compared on the same Max-Cut instance ($n=10$ and $P=7)$, where SSR-MCTS obtains a better approximation ratio. A comparison of the respective resource costs, however, is less straightforward as our algorithm leverages the knowledge of optimal schedules on shallower circuits and, therefore, it depends on the details of the implementation of the iterative procedure.

A resource-savvy implementation will invest more time in the optimization of low-depth ans\"atze and progressively reduce the number of cycles per layer as $P$ increases up to the desired accuracy or until the energy no longer decreases.
In cases where the number of function evaluations is an important limiting factor of a particular experimental implementation of QAOA, the branching factor could also be reduced to make the algorithm converge with fewer calls at the cost of the reachability of the SSR-MCTS algorithm. These coarser results could also be used as good initial parameters for gradient-based runs (see Appendix~\ref{app:globalopt}). Additionally, integrating MCTS with neural networks, as done in Ref.~\cite{chen_mctsQA} for annealing schedules, will likely greatly enhance our strategy. Indeed, the neural networks allow for adapting existing trained MCTS agents on new problem instances or circuit depths without having to learn the decision tree structure from scratch.

A separate discussion concerns the number of shots needed for any function evaluation. To estimate the reward with an accuracy $\epsilon$, typically one needs $N_{\rm shots} = O(1/\epsilon^2)$ single-shot measurements.
Here, SSR-MCTS has two nice features that contribute towards reducing this overhead. 
First, the robustness against measurement noise suggests that rough estimates of the reward are sufficient for obtaining good approximations in the ground states, in particular in the early training stage of the MCTS agent. 
Second, The prefactor in the proportionality relation between $N_{\rm shots}$ and $\epsilon^{-2}$ will likely decrease as $P$ increases. Indeed, SSR already severely restricts the energy window spanned by the QAOA ansatz, as clearly shown by Fig.~\ref{fig:leaf_distr}. Their energy variance is reduced accordingly.
Finally, one can train the MCTS algorithm directly with single-shot measurements and update the value of the tree nodes using the energy of the sampled classical configuration.
This might greatly reduce the resource requirements but further investigations are needed to gauge its efficacy and possible advantage over the approach presented here.
\\

\providecommand{\noopsort}[1]{}\providecommand{\singleletter}[1]{#1}%

\end{document}